\documentclass[preprint]{aastex}
\usepackage{graphicx}
\pdfoutput=1

\shorttitle{Three-dimensional propagation of magnetohydrodynamic waves}
\shortauthors{Rial et al.}

\begin{document}

\title{THREE-DIMENSIONAL PROPAGATION OF MAGNETOHYDRODYNAMIC WAVES IN SOLAR CORONAL ARCADES}
\author{S. Rial\altaffilmark{1}, I. Arregui\altaffilmark{1}, J. Terradas\altaffilmark{1,2}, R. Oliver\altaffilmark{1} and J. L. Ballester\altaffilmark{1}}

\altaffiltext{1}{Departament de F\'{\i}sica, Universitat de les Illes Balears,
E-07122 Palma de Mallorca, Spain. Email: samuel.rial@uib.es, inigo.arregui@uib.es,
jaume.terradas@uib.es, ramon.oliver@uib.es and joseluis.ballester@uib.es}
\altaffiltext{2}{Centrum voor Plasma Astrofysica, K.U. Leuven,
Celestijnenlaan 200B, B-3001 Heverlee, Belgium}

\date{ }

\begin{abstract}
We numerically investigate the excitation and temporal evolution of oscillations in a two-dimensional coronal arcade by including the three-dimensional propagation of perturbations. The time evolution of impulsively generated  perturbations is studied by solving the linear, ideal magnetohydrodynamic (MHD) equations in the zero-$\beta$ approximation. As we neglect gas pressure the slow mode is absent and therefore only coupled MHD fast and Alfv\'en modes remain. Two types of numerical experiments are performed. First, the resonant wave energy transfer between a fast normal mode of the system and local Alfv\'en waves is analyzed. It is seen how, because of resonant coupling, the fast wave with global character transfers its energy to Alfv\'enic oscillations localized around a particular magnetic surface within the arcade, thus producing the damping of the initial fast MHD mode. Second, the time evolution of a localized impulsive excitation, trying to mimic a nearby coronal disturbance, is considered. In this case, the generated fast wavefront leaves its energy on several magnetic surfaces within the arcade. The system is therefore able to trap energy in the form of Alfv\'enic oscillations, even in the absence of a density enhancement such as that of a coronal loop. These local oscillations are subsequently phase-mixed to smaller spatial scales. The amount of wave energy trapped by the system via wave energy conversion strongly depends on the wavelength of perturbations in the perpendicular direction, but is almost independent from the ratio of the magnetic to density scale heights.
\end{abstract}


\section{Introduction}
\label{Introduction}
The presence of waves and oscillations in the solar corona is a well known feature that has been observed for long time. For an overview of the early observational background see \citet{T1988}. Nowadays, because of the increasing spatial and temporal resolution of the EUV instruments onboard TRACE, SOHO and HINODE spacecraft, accurate observations of oscillations in different coronal structures are accomplished. Many authors have reported observations of transversal coronal loop oscillations from both ground and space-based instruments \citep{AFS1999,NOD1999,APS2002,SAT2002}. When these observations are compared with theoretical models  \citep{REB1984,NOD1999,NO2001}, the possibility of inferring some plasma parameters, otherwise difficult to measure, and of improving the existing theoretical models is open; see \citet{BEO2007} for a review. Magnetohydrodynamics (MHD) is the underlying theory of coronal seismology and it is believed that all these observed oscillations and waves can be interpreted theoretically in terms of MHD modes of different coronal plasma structures. 

The theoretical study of these oscillations and waves can be done from several points of view. The first approach is to make a normal mode analysis of the linearized MHD equations, which allows to obtain the spatial distribution of the eigenmodes of the structure together with the dispersion relation $\omega(\bf{k})$. Once the elementary building blocks of the MHD normal mode theory are described, the main properties of the resulting MHD waves can be outlined. Many authors have explored the normal modes of coronal structures, beginning with very simple cases such as the straight and infinite cylinder \citep{ER1983}. In the context of curved coronal magnetic structures, \citet{GPH1985,PHG1985,PG1988} investigated the continuous spectrum of ideal MHD. \citet{OBH1993,OHP1996} and \citet{TOB1999} derived the spectrum of modes in potential and nonpotential arcades. More complex configurations, such as sheared magnetic arcades in the zero-$\beta$ plasma limit, have been studied by \citet{AOB2004,AOB2004a}. Other authors have studied eigenmodes in curved configurations with density enhancements that represent coronal loops \citep[e.g.,][]{VDA2004,TOB2006,VFN2006a,VFN2006b,VFN2006c,DAR2006,VVT2009}. 

An alternative approach is to obtain the time dependent solution of the MHD equations. Using this method, \citet{CB1995,CB1995a} studied analytically the propagation of fast waves in a two-dimensional coronal arcade for a particular equilibrium, namely one with uniform Alfv\'en speed. \citet{OMB1998} studied the effect of impulsively generated fast waves in the same coronal structure. \citet{DSV2005} studied the properties of Alfv\'en waves in an arcade configuration, including the transition region between the photosphere and the corona. Other studies have analyzed the effect of the loop structure on the properties of fast and slow waves in two-dimensional curved configurations \citep[see, e.g.,][]{MSN2005,BA2005,BVA2006,SSM2006,SMS2007}, see \citet{T2009} for a review. 

The main aim of this paper is to analyze the effect of including three-dimensional propagation on the resulting MHD waves as a first step before considering more realistic situations like the one observed by \citet{VNO2004}, where the effect of three-dimensional propagation is clear. In our model there is no density enhancement like that of a loop and the zero-$\beta$ approximation is assumed, so only the fast and Alfv\'en modes are present. We focus our attention on the mixed properties displayed by the generated MHD waves that arise due to the coupling when longitudinal propagation is allowed. The paper is arranged as follows. In \S~\ref{equilibrium_conf} we briefly describe the equilibrium configuration as well as some of the approximations made in this work. In \S~\ref{linear} we present our derivation of the linear ideal MHD wave equations with three-dimensional propagation of perturbations. In \S~\ref{numerical_method_and_test} the numerical code used in our study is described, together with several checks that have been performed by solving problems with known analytical or simple numerical solution. Our main results are shown in \S~\ref{numerical_res}, where the linear wave propagation properties of coupled fast and Alfv\'en waves in a two-dimensional coronal arcade, allowing three-dimensional propagation, are described. Finally, in \S~\ref{conclusions} the conclusions are drawn.

\section{Equilibrium configuration}  
\label{equilibrium_conf}
\begin{figure}[t]
\begin{center}
\hspace{-0.5cm}\includegraphics[width=0.7\textwidth]{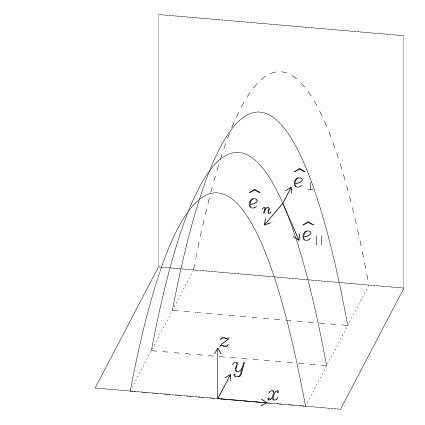}
\caption{Sketch of the magnetostatic configuration of a potential coronal arcade, where the solid curves represent magnetic field lines, given by $A(x,z)=$ constant. These curves in the $xz$-plane become arcade surfaces in three dimensions. In this model $z$ measures the upward distance from the base of the corona (placed at $z=0$). The three orthogonal unit vectors, $\mathbf{\hat{e}}_{n}$, $\mathbf{\hat{e}}_{\bot}$ and  $\mathbf{\hat{e}}_{\Vert}$, defining the normal, perpendicular and parallel directions respectively, are also shown at a particular point.}
\label{fig:arc}
\end{center}
\end{figure}

We model a solar coronal arcade by means of a two-dimensional potential configuration contained in the $xz$-plane in a Cartesian system of coordinates \citep[see][]{OBH1993}. For this $y$-invariant configuration the flux function is

\begin{equation} 
A(x,z)=B\Lambda_{B}\cos{\left(\frac{x}{\Lambda_{B}}\right)}\exp{\left(-\frac{z}{\Lambda_{B}}\right)},
\label{eq:flux}
\end{equation}
 and the magnetic field components are given by

\begin{displaymath} 
B_{x}(x,z)=B\cos\left(\frac{x}{\Lambda_{B}}\right)\exp\left({-\frac{z}{\Lambda_{B}}}\right),
\end{displaymath}
\begin{equation}
B_{z}(x,z)=-B\sin\left(\frac{x}{\Lambda_{B}}\right)\exp\left({-\frac{z}{\Lambda_{B}}}\right).
\label{eq:arccomp}
\end{equation}
In these expressions $\Lambda_{B}$ is the magnetic scale height, which is related to the lateral extent of the arcade, $L$, by $\Lambda_{B}=2L/\pi$, and $B$ represents the magnetic field strength at the photospheric level ($z=0$). The overall shape of the arcade is shown in Figure~\ref{fig:arc}. 

In this paper gravity is neglected and the $\beta=0$ approximation is used for simplicity. Therefore, the equilibrium density can be chosen arbitrarily. We adopt the following one-dimensional profile

\begin{equation} 
\rho_{0}(z)=\rho_{0}\exp\left({-\frac{z}{\Lambda}}\right),
\label{eq:density}
\end{equation}
where $\Lambda$ is the density scale height and $\rho_{0}$ is the density at the base of the corona. As shown by \citet{OBH1993}, the combination of magnetic field components given by Equation~(\ref{eq:arccomp}) with the density profile given by Equation~(\ref{eq:density}) leads to a one-dimensional Alfv\'en speed distribution in the arcade that can be cast as 

\begin{equation} 
v_{A}(z)=v_{A0}\exp{\left[-(2-\delta)\frac{z}{2\Lambda_{B}}\right]}.
\label{eq:Alfven1}
\end{equation}
Here $\delta=\frac{\Lambda_{B}}{\Lambda}$ represents the ratio of the magnetic scale height to the density scale height and $v_{A0}$ is the Alfv\'en speed at the base of the corona. The $\delta$ parameter completely determines the behavior of the Alfv\'en speed profile and hence the wave propagation properties. The case $\delta=2$ represents a uniform Alfv\'en speed model, while $\delta=0$ corresponds to an exponentially decreasing Alfv\'en speed in a uniform density configuration. Other values of $\delta$ represent situations in which both the Alfv\'en speed and density depend on height in a different manner.

\section{Linear waves}
\label{linear}
 In order to study small amplitude oscillations in our potential arcade the previous equilibrium is perturbed. For linear and adiabatic MHD perturbations in the zero-$\beta$ approximation the relevant equations are 
 
\begin{equation} 
\rho_{0}\frac{\partial\mathbf{v}_{1}}{\partial t}=\frac{1}{\mu_{0}}(\nabla\times\mathbf{B})\times\mathbf{B}_{1}+\frac{1}{\mu_{0}}(\nabla\times\mathbf{B}_{1})\times\mathbf{B},
\label{eq:momentum}
\end{equation}

\begin{equation} 
\frac{\partial \mathbf{B}_{1}}{\partial t} =\nabla\times(\mathbf{v}_{1}\times\mathbf{B}),
\label{eq:induction}
\end{equation}
where $\mu_{0}$ is the magnetic permeability of free space and the subscript ``$1$'' is used to represent perturbed quantities. These equations are next particularized to our two-dimensional potential arcade equilibrium. As the equilibrium is invariant in the $y$-direction, we can Fourier analyze all perturbed quantities in the $y$-direction by making them proportional to $\exp{(ik_{y}y)}$. In this way, three-dimensional propagation is allowed and each Fourier component can be studied separately. As a result of this Fourier analysis the perpendicular perturbed velocity and magnetic field components appear accompanied by the purely imaginary number $i=\sqrt{-1}$. This is undesirable from a practical point of view, since Equations~(\ref{eq:momentum}) and (\ref{eq:induction}) will be solved numerically and the code is designed to handle real quantities only. Nevertheless, by making the appropriate redefinitions, namely $v_{1y}\equiv i\tilde{v}_{1y}$ and  $B_{1y}\equiv i\tilde{B}_{1y}$, it turns out that our wave equations can be cast in the following form

\begin{eqnarray}
\frac{\partial v_{1x}}{\partial t}&=&\frac{1}{\mu_{0}\rho_{0}}\Bigg[\left(\frac{\partial B_{1x}}{\partial z}-\frac{\partial B_{1z}}{\partial x}\right)B_{z}\Bigg],\label{eq:velocityx}\\
\frac{\partial \tilde{v}_{1y}}{\partial t}&=&\frac{1}{\mu_{0}\rho_{0}}\Bigg[\left(B_{x}\frac{\partial\tilde{B}_{1y}}{\partial x}+B_{z}\frac{\partial\tilde{B}_{1y}}{\partial z}\right)+k_{y}\left(B_{1x}B_{x}+B_{1z}B_{z}\right)\Bigg],\label{eq:velocityy}\\
\frac{\partial v_{1z}}{\partial t}&=&-\frac{1}{\mu_{0}\rho_{0}}\Bigg[\left(\frac{\partial B_{1x}}{\partial z}-\frac{\partial B_{1z}}{\partial x}\right)B_{x}\Bigg],\label{eq:velocityz}\\
\frac{\partial B_{1x}}{\partial t}&=&-k_{y}\tilde{v}_{1y}B_{x}-\frac{\partial}{\partial z}\left(v_{1z}B_{x}-v_{1x}B_{z}\right), \label{eq:fieldx}\\
\frac{\partial \tilde{B}_{1y}}{\partial t}&=&\frac{\partial}{\partial z}\left(\tilde{v}_{1y}B_{z}\right)+\frac{\partial}{\partial x}\left(\tilde{v}_{1y}B_{x}\right),\label{eq:fieldy}\\
\frac{\partial B_{1z}}{\partial t}&=&-k_{y}\tilde{v}_{1y}B_{z}+\frac{\partial}{\partial x}\left(v_{1z}B_{x}-v_{1x}B_{z}\right).\label{eq:fieldz}\\
\nonumber
\end{eqnarray}

These equations constitute a set of coupled partial differential equations with non-constant coefficients that describe the propagation of fast and Alfv\'en waves. As the plasma $\beta=0$, slow waves are excluded from the analysis. When $k_{y}=0$, equations~(\ref{eq:velocityx})--(\ref{eq:fieldz}) constitute two independent sets of equations. The two equations for $\tilde{v}_{1y}$ and $\tilde{B}_{1y}$ are associated to Alfv\'en wave propagation. On the other hand, the four equations for the remaining variables, $v_{1x}$, $v_{1z}$, $B_{1x}$, $B_{1z}$, describe the fast wave propagation. The basic normal mode properties of fast and Alfv\'en modes in a potential arcade with $k_{y}=0$ are described in \citet{OBH1993}, while the case $k_{y}\neq0$ has been considered by \citet{AOB2004}. The time dependent propagation for $k_{y}=0$ was analyzed by \citet{TOB2008}. When longitudinal propagation of perturbations is allowed ($k_{y}\neq0$), the six equations and their solutions are coupled so we may anticipate fast and Alfv\'en wave propagation to display a mixed nature, in an analogous way to the mixed character of eigenmodes obtained by \citet{AOB2004} in their analysis of the normal modes of the present equilibrium with $k_{y}\neq0$. In the following the tildes in $\tilde{v}_{1y}$ and $\tilde{B}_{1y}$ are dropped.

\section{Numerical method and test cases}
\label{numerical_method_and_test}
\subsection{Numerical method}
\label{numerical_method}
The set of differential equations (\ref{eq:velocityx})--(\ref{eq:fieldz}) is too complicated to have analytical or simple numerical solutions except for simplified configurations and under particular assumptions. For this reason we solve them by using a numerical code, although comparisons with known wave properties have been carried out whenever possible. 

When considering a potential arcade as the equilibrium magnetic field, it is advantageous to use field-related components instead of Cartesian components in order to characterize the directions of interest related to the polarization of each wave type. The unit vectors in the directions normal, perpendicular, and parallel to the equilibrium magnetic field are given by

\begin{equation} 
\mathbf{\hat{e}}_{n}=\frac{\nabla A}{\mid\nabla A\mid},\quad\mathbf{\hat{e}}_{\bot}=\mathbf{\hat{e}}_{y},\quad\mathbf{\hat{e}}_{\Vert}=\frac{\mathbf{B}}{\mid\mathbf{B}\mid},
\label{eq:genunitvec}
\end{equation}
where $A$ is the flux function given in Equation~(\ref{eq:flux}). These unit vectors are related to the Cartesian ones as follows

\begin{equation} 
\mathbf{\hat{e}}_{n}=\frac{(B_{z}\mathbf{\hat{e}}_{x}-B_{x}\mathbf{\hat{e}}_{z})}{\mid\mathbf{B}\mid},\quad\mathbf{\hat{e}}_{\bot}=\mathbf{\hat{e}}_{y},\quad\mathbf{\hat{e}}_{\Vert}=\frac{(B_{x}\mathbf{\hat{e}}_{x}+B_{z}\mathbf{\hat{e}}_{z})}{\mid \mathbf{B}\mid},
\label{eq:potentialunitvec}
\end{equation}
 with $\mid \mathbf{B}\mid=(B_{x}^2+B_{z}^2)^{1/2}$. In the absence of longitudinal propagation (i.e. for $k_{y}=0$), these three directions are associated with the three types of waves that can be excited, namely $v_{1n}$ for fast waves, $v_{1\bot}$ for Alfv\'en waves, and $v_{1\Vert}$ for slow waves. 
 
 Since we want to model a coronal disturbance with a localized spatial distribution we have considered as the initial condition a two-dimensional Gaussian profile given by

\begin{equation} 
v_{1}=v_{s}\exp\left[-\frac{(x-x_{s})^2+(z-z_{s})^2}{a^2}\right],
\label{eq:perturbation}
\end{equation}
where $v_{s}$ is the amplitude of the velocity perturbation, $x_{s}$ and $z_{s}$ are the coordinates of the perturbation's center, and $a$ is the width of the Gaussian profile at half height. In the following we use $v_{1}=v_{1n}$ to excite fast waves and $v_{1}=v_{1y}$ to excite Alfv\'en waves. When $k_{y}=0$ the fast mode produces plasma motions purely normal to the magnetic field, while the Alfv\'en mode is characterized by a purely perpendicular velocity component. When propagation along the $y$-direction is considered, pure fast or Alfv\'en modes do not exist and both produce motions in the normal velocity component as well as in the perpendicular velocity component \citep{AOB2004}. It must be noted that the numerical code solves the time-dependent equations in Cartesian coordinates and so the solution has to be transformed following expressions (\ref{eq:potentialunitvec}) to the field-related coordinates. The same applies to the initial perturbation, which must be transformed into the corresponding Cartesian components. 

The numerical code \citep[see][for details about the method]{BBT2009} uses the so-called method of lines for the discretization of the variables and the time and space variables are treated separately. For the temporal part, a fourth-order Runge-Kutta method is used. For the space discretization a finite-difference method with a fourth-order centered stencil is choosen. For a given spatial resolution, the time step is selected so as to satisfy the Courant condition. As for the boundary conditions, as we computed the time evolution of two initial perturbations, two kinds of boundary conditions are used. First, when the initial perturbation in $v_{1n}$ is the fundamental normal mode of the $k_{y}=0$ problem, for the $v_{1n}$ component line-tying conditions are chosen at all boundaries, while for the $v_{1y}$ component, flow-through conditions are selected except at $z=0$ where line-tying condition is used. On the other hand, when an initial perturbation like (\ref{eq:perturbation}) is considered, the large photospheric inertia is accomplished by imposing line-tying boundary conditions at $z=0$. In all other boundaries flow-through conditions are used so that perturbations are free to leave the system. In order to increase numerical stability, fourth-order artificial dissipation terms are included in the numerical scheme. In all the simulations the effects of this artificial dissipation have been checked to ensure that they do not affect the obtained solution, but just contribute to eliminate undesired high-frequency numerical modes.

\subsection{Test cases}
\label{test_cases}
Some preliminary tests have been performed in order to figure out the appropriate values of numerical parameters, such as the grid resolution or the numerical dissipation, on the obtained results for fast and Alfv\'en waves. The first test we have conducted has been to run the code with no perturbation at all and to check that the structure remains stable. The results of this numerical run were completely satisfactory. Then the propagation of linear fast and Alfv\'en MHD waves in a potential coronal arcade has been considered. 

\subsubsection{Fast wave}
\label{fast_wave}
The temporal evolution of impulsively generated perturbations with rather similar conditions has been accomplished by several authors: \citet{CB1995} obtained analytical expressions for the temporal evolution of perturbations when a coronal arcade is taken as the equilibrium state; \citet{CB1995a, OMB1998} numerically computed such solutions when different initial perturbations are used; and more recently \citet{TOB2008} showed the main properties of the time evolution of fast and Alfv\'en waves in low-$\beta$ environments. These works facilitate the comparison of our numerical results with known results as well as with analytical ones. 

As shown by \citet{TOB2008}, when different resolutions are used time dependent results reveal that the grid resolution in our two-dimensional domain is not a critical factor for the proper computation of fast waves and that a good representation of the temporal evolution of perturbations can be achieved even with a rather modest resolution of $50\times50$ grid points in the $(x,z)$-plane. As mentioned above, numerical dissipation is introduced in our code in order to ensure numerical stability. This dissipation is proportional to an adjustable parameter, or dissipation factor, $\sigma_{n}$. We have conducted numerical simulations for different values of the dissipation factor and it turns out that the temporal evolution of fast wave perturbations is not modified.

\subsubsection{ Alfv\'en wave}
\label{alfven_wave}
The properties of Alfv\'en continuum normal modes in a potential coronal arcade described by \citet{OBH1993} allow us to anticipate and identify possible sources of difficulties in the numerical computation of Alfv\'en wave solutions. First of all, since they are oscillatory solutions strongly confined around given magnetic surfaces (both when propagating or in their standing mode version), spatial scales quickly decrease with time and so we can expect a rather important dependence of the numerical solutions on the number of grid points used to cover the area in and around the excited magnetic surfaces. The situation becomes even worse if we take into account that computations in a Cartesian grid do not allow us to locate all the grid points along magnetic surfaces. This fact affects the numerical results and adds a numerical damping. Furthermore, when time-dependent simulations are considered the sampling rate is no more an independent parameter. When the spatial resolution of the grid is defined, the Courant condition gives a maximum value for the temporal resolution which in turn sets the maximum frequency that can be resolved.

\begin{figure}[!t]
\begin{center}
\includegraphics[width=0.42\textwidth]{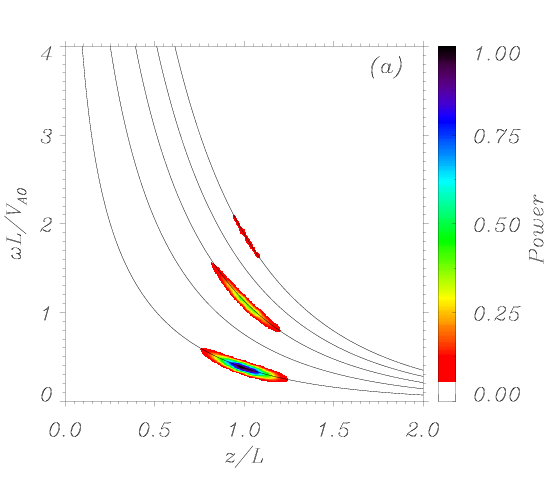}
\includegraphics[width=0.49\textwidth]{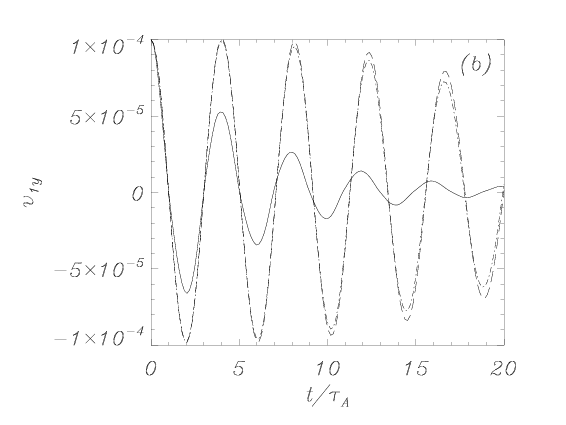}
\caption{(a) Shaded contours are the power spectrum of $v_{1y}$ at $x=0$ as a function of height, $z/L$, and normalized frequency, $\omega L/v_{A0}$. In this simulation the spatial grid is set to $600\times600$ while the numerical dissipation is fixed to $\sigma_{n}=0.001$. Solid lines are the theoretical frequency of the normal Alfv\'en modes for $\delta=0$, given by \citet{OBH1993}. From bottom to top they represent the fundamental mode and its harmonics.\ (b) Temporal evolution of the $v_{1y}$ component when the initial perturbation is an Alfv\'en normal mode of the system located at $x=0$ and $z_{m}/L=0.33$ (see Equation~[\ref{eq:normalpert}]). Here $\delta=0$, the numerical grid has $400\times400$ points, and $\sigma_{n}=0.001$. Different solutions correspond to $a/L=0.01$ (solid), $a/L=0.1$ (dash-dotted) and $a/L=0.2$ (long-dashed). Time is given in units of $\tau_{A}=L/v_{A0}$.}
\label{fig:spectrumalfvenky0}
\end{center}
\end{figure}

We have first generated Alfv\'en waves in our potential arcade model by considering an impulsive initial excitation of the $v_{1y}$ component given by Equation~(\ref{eq:perturbation}) with $z_{s}=1$ and $x_{s}=0$. This implies that the initial disturbance is even about $x=0$ and so odd Alfv\'en modes are not excited. As described by \citet{TOB2008}, the spatial resolution of the numerical mesh affects the obtained amplitude and frequency values. Better resolution provides a closer value to the analytical frequency and less numerical damping. We have also checked the influence of numerical dissipation and the results show that only the amplitude, and therefore the damping time, decreases when the $\sigma_{n}$ parameter is decreased. The spectral analysis of these oscillations at different heights in the structure is shown in Figure~\ref{fig:spectrumalfvenky0}a. The resulting power spectrum is compared to the Alfv\'en continuum frequencies obtained by \citet{OBH1993}. The frequency associated with the generated Alfv\'en waves coincides with the theoretical normal mode frequencies of the system, which gives us further confidence on the goodness of our code. Alfv\'en waves stay confined to the vertical range of magnetic surfaces that were excited by the initial disturbance, since they cannot propagate energy across magnetic surfaces. The initial perturbation is decomposed by the system in a linear combination of normal modes, but keeping the even parity of the initial disturbance with respect to $x=0$, so energy is only found in the fundamental mode, the second harmonic, etc.

In order to better isolate and show the possible numerical artifacts that the code introduces into the numerical solution we have considered a simpler case, the excitation of a particular Alfv\'en mode around a magnetic surface. According to \citet{OBH1993}, Alfv\'en normal mode solutions can be obtained analytically when $\delta=0$. For this reason we now select $\delta=0$.

The initial excitation could now be given by

\begin{equation}
v_{1y}(x,z)=\hat{v}_{1y}(x)\delta\Big[A(x,z)-A(x=0,z_{m})\Big],
\label{eq:normalmode}
\end{equation}
where $\hat{v}_{1y}$ is the regular part of the solution, $A(x,z)$ is the flux function defined by Equation~(\ref{eq:flux}), and $z_{m}$ gives the maximum height of the magnetic field line in which the normal mode is excited. It is important to note that the regular solution has a well-defined parity with respect to the $x$-direction depending on whether $n$ is chosen even or odd. However, since a delta function is difficult to handle from a numerical point of view, our normal mode-like excitation is performed by an initial perturbation of the form

\begin{equation}
v_{1y}(x,z)=\hat{v}_{1y}(x)\exp{\left[-\frac{A(x,z)-A(x=0,z_{m})}{a^2}\right]}.
\label{eq:normalpert}
\end{equation}
For the regular part, $\hat{v}_{1y}(x)$, the fundamental mode with one maximum along the field lines has been chosen. It should be noted that the width, $a$, of the initial perturbation now causes the excitation of several Alfv\'en modes in a set of neighboring magnetic surfaces. It is important to consider an initial velocity profile which is sufficiently localized in the direction transverse to magnetic surfaces so that only a few of them are excited. As we concentrate on the dynamics of a restricted number of field lines around a magnetic surface the consideration of other models, with different values of $\delta$, would change quantitatively the generated frequencies, but not the overall qualitative conclusions shown here.

Figure \ref{fig:spectrumalfvenky0}b shows the temporal evolution of the excited $v_{1y}$ component at a particular location as a function of time for three different values for the width of the initial disturbance. It is clear that three different solutions are obtained. The two corresponding to the largest widths are rather similar, but the one for the smaller width shows a strong damping. It must be said that the exact solution of this ideal system should display no time damping, hence we assert that this is a numerical effect, that cannot be attributed to a real physical damping mechanism. This undesired effect is less important for larger widths of the initial perturbation since, for a given number of grid points, the initial condition is better resolved spatially.

\begin{figure}[!t]
\begin{center}
\includegraphics[width=0.46\textwidth]{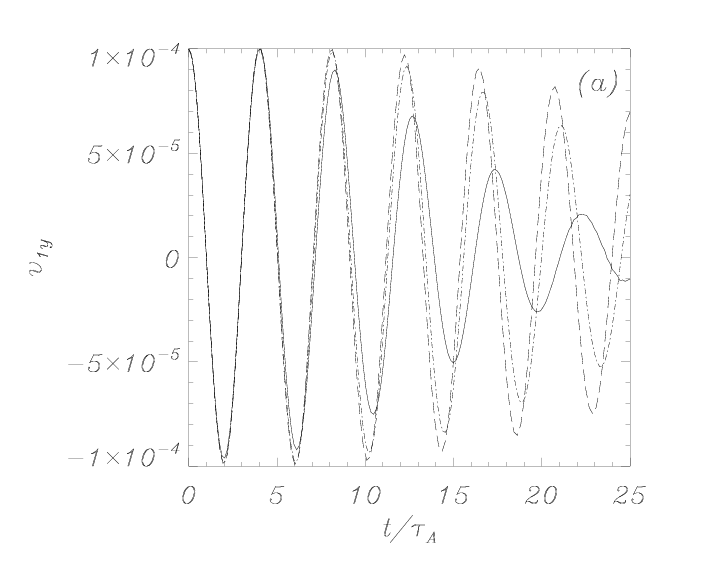}
\includegraphics[width=0.46\textwidth]{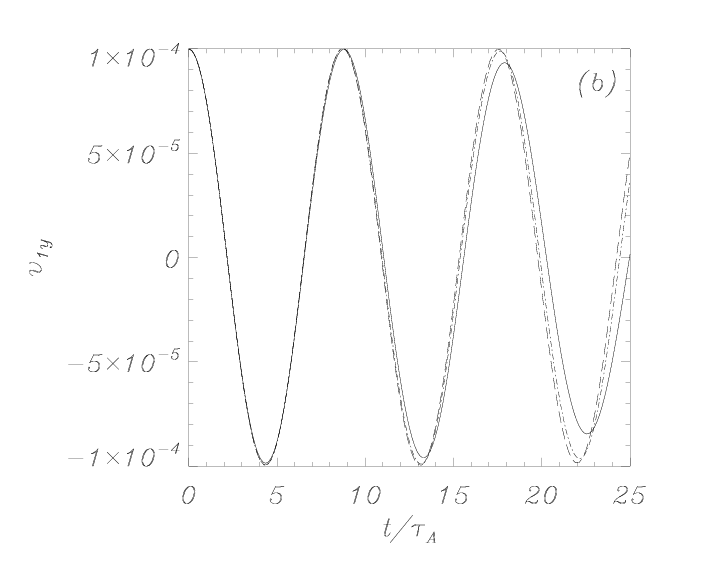}
\caption{(a) Temporal evolution of the $v_{1y}$ velocity component at $x=0$ and $z_{m}/L=0.33$ for an initial perturbation given by Equation~(\ref{eq:normalpert}) with $a/L=0.2$ and $\sigma_{n}=0.001$. \ (b) Temporal evolution of the $v_{1y}$ velocity component at $x=0$ and $z_{m}/L=0.66$ for an initial perturbation with $a/L=0.2$ and $\sigma_{n}=0.001$. In both panels solid, dash-dotted, and long-dashed lines represent a resolution of $200\times200$, $400\times400$, and $600\times600$ points, respectively.}
\label{fig:resolution}
\end{center}
\end{figure}

We next fix the width of the initial disturbance, $a$, and vary the spatial resolution in our domain. Figure \ref{fig:resolution} shows several numerical simulations when an initial normal mode-like excitation (Equation~[\ref{eq:normalpert}]) is made at different heights. It is clear that larger spatial resolution provides the more accurately the undamped oscillatory solution. 

Also from this analysis we conclude that the spatial resolution is not a factor that should be taken into account in an isolated manner when considering the numerical description of Alfv\'en waves on given magnetic surfaces. Indeed, and because of the Cartesian distribution of grid points in a system of curved magnetic field lines, low-lying magnetic lines are poorly resolved when compared to high-lying magnetic lines for a given grid resolution. This has implications that are worth to be taken into account as can be seen in Figure~\ref{fig:resolution}. If we compare signals in Figure~\ref{fig:resolution}, we can see that, all parameters being the same, closer results to the analytical solution are obtained for higher magnetic field lines. We can therefore assert that for the numerical simulation of Alfv\'en wave properties the resolution of the grid is an important parameter and that it becomes more critical for low-lying magnetic field lines than for higher ones. It should be noted that the conclusions of these tests can also be applied to the case in which an impulsive excitation is set as the initial perturbation.

\section{Numerical results}
\label{numerical_res}
In this section we present the main results from our numerical investigation. For simplicity, first, the temporal evolution of a normal mode-like fast disturbance is analyzed in order to show how and where resonant absorption, due to three-dimensional propagation of perturbations in a non-uniform medium, takes place. It turns out that previous results obtained for the normal modes of coupled fast and Alfv\'en waves in a potential arcade by \citet{AOB2004} can guide us to understand the time evolution of the system and the energy transfer between resonantly coupled modes. Then, a more complex situation is considered by analyzing the time evolution of the initial perturbation given by Equation~(\ref{eq:perturbation}). It should be noted that our first normal mode time evolution analysis has been proof very useful to further better understand the resulting coupling process between both velocity components when a localized impulsive disturbance is used.

\subsection{Resonant damping of fast MHD normal modes in a potential arcade}
\label{normal_like}
In order to gain some insight into the propagation properties of coupled fast and Alfv\'en waves in our configuration, we first study the time evolution caused by an initial disturbance having the spatial structure of a fast normal mode for $k_{y}=0$ (propagation in the $xz$-plane). As shown by \citet{OBH1993}, pure fast modes in a potential arcade are characterized by a global spatial structure determined by the wavenumbers $k_{x}$ and $k_{z}$, which give rise to smooth distributions with a given number of maxima in the $x$- and $z$-directions. This results in a discrete spectrum of frequencies. The frequencies and spatial structure of the fast modes with $k_{y}\neq0$ were computed by \citet{AOB2004}, who showed that perpendicular propagation produces the coupling of the fast normal modes to Alfv\'en continuum solutions, resulting in modes with mixed properties. We have chosen as initial perturbation the velocity perturbation $v_{1n}$ of the fundamental fast mode for $k_{y}=0$, with one maximum in each direction in the $xz$-plane. When $k_{y}=0$ this produces a standing harmonic oscillation of the system, as in an elastic membrane. When $k_{y}\neq0$, this initial perturbation is not a normal mode of the system, but we expect that, the obtained temporal evolution will not differ very much from the actual normal mode of the coupled solution.

\begin{figure*}[!t]
\begin{center}
\hspace{-0.3cm}\includegraphics[width=0.31\textwidth]{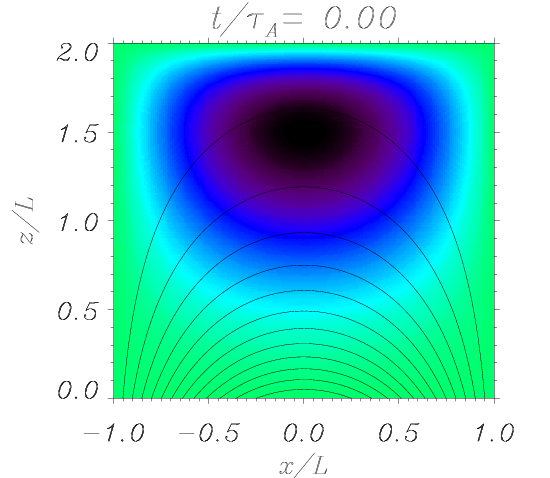}
\includegraphics[width=0.31\textwidth]{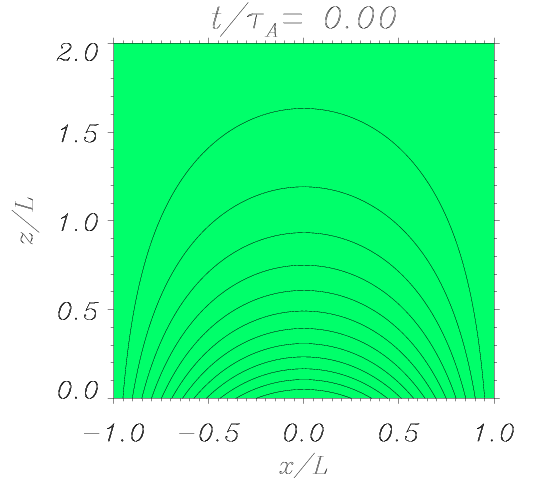}
\includegraphics[width=0.31\textwidth]{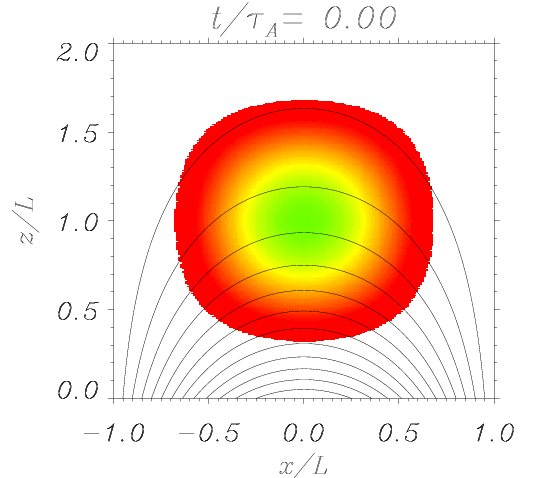}\\
\hspace{-0.3cm}\includegraphics[width=0.31\textwidth]{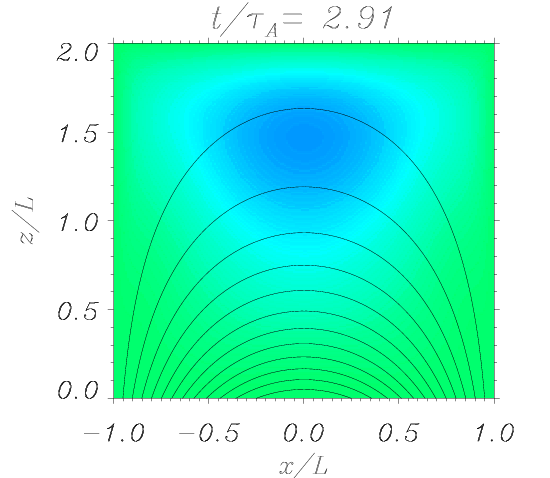}
\includegraphics[width=0.31\textwidth]{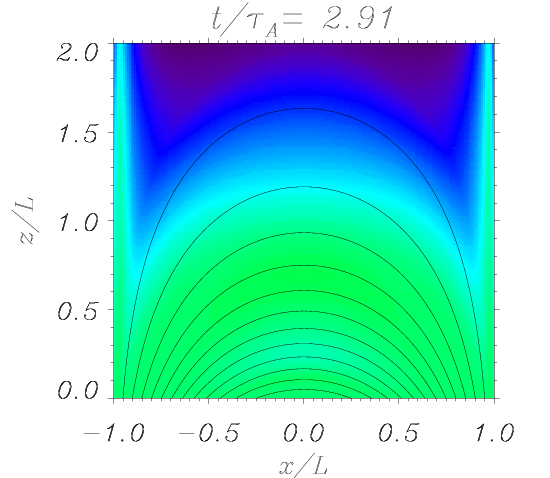}
\includegraphics[width=0.31\textwidth]{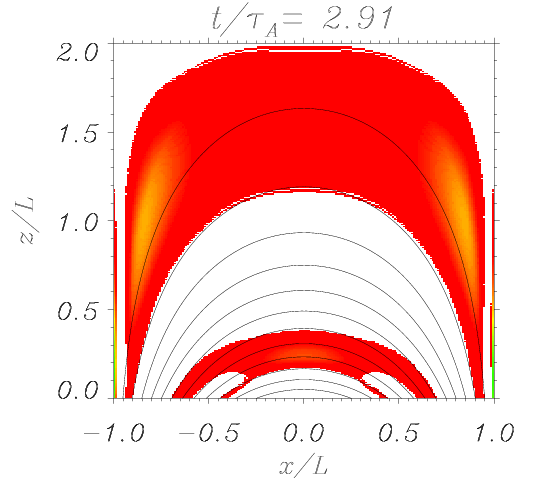}\\
\hspace{-0.3cm}\includegraphics[width=0.31\textwidth]{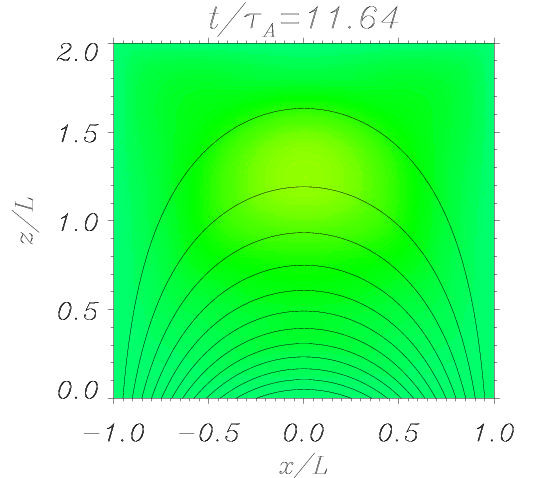}
\includegraphics[width=0.31\textwidth]{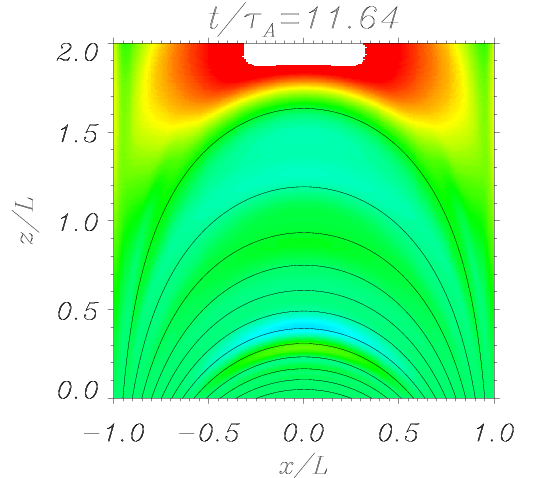}
\includegraphics[width=0.31\textwidth]{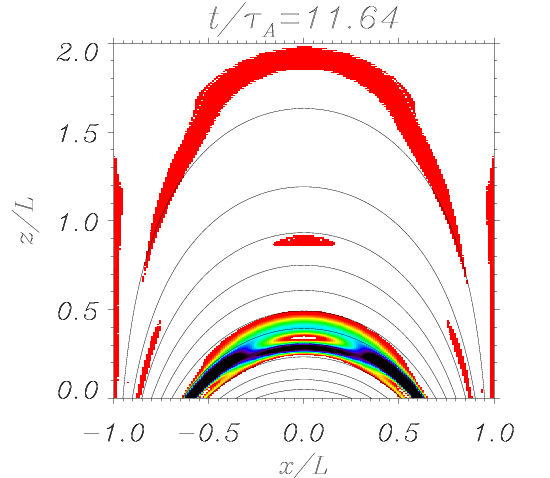}\\
\hspace{-0.3cm}\includegraphics[width=0.31\textwidth]{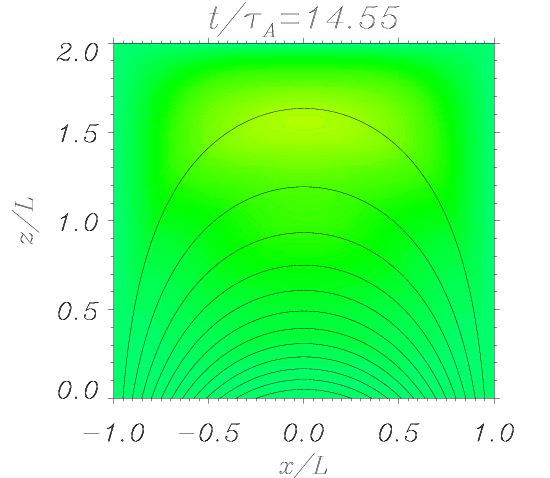}
\includegraphics[width=0.31\textwidth]{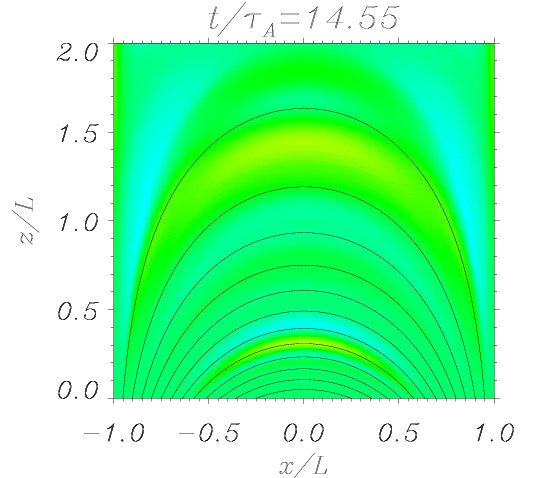}
\includegraphics[width=0.31\textwidth]{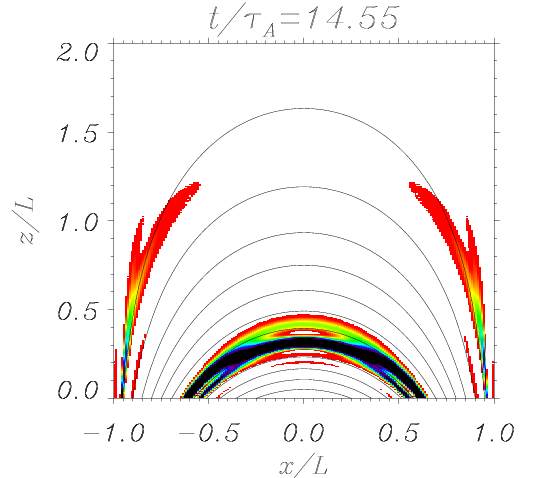}\\
\hspace{0.4cm}\includegraphics[width=0.31\textwidth]{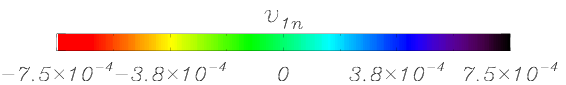}
\includegraphics[width=0.31\textwidth]{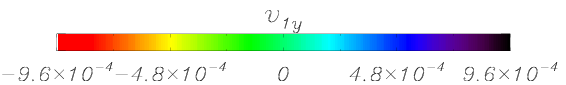}
\hspace{-0.1cm}\includegraphics[width=0.31\textwidth]{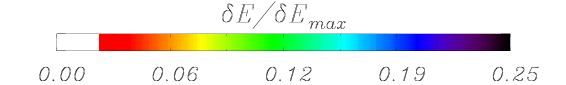}
\caption{Several snapshots of the two velocity components, $v_{1n}$ and $v_{1y}$, and of the total energy density in a potential arcade with $\delta=2$ and for $k_{y}L=1$. The initial perturbation in $v_{1n}$ is the fundamental normal mode of the $k_{y}=0$ problem. For this simulation a $600\times600$ grid is used. Magnetic field lines are represented with white (left and middle panels) and black lines (right panels). A movie displaying the full time evolution is available in the electronic version of the journal.}
\label{fig:chap4nomarlkyn0}
\end{center}
\end{figure*}

\begin{figure}[!h]
\begin{center}
\includegraphics[width=\textwidth]{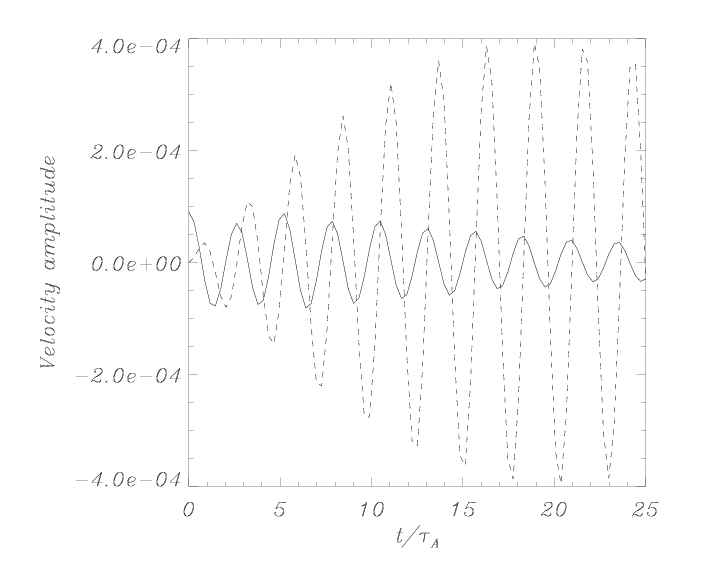}
\caption{Temporal evolution of the normal, $v_{1n}$ (solid line), and perpendicular, $v_{1y}$ (dashed line), velocity components at $x=0$, $z/L=0.35$. Data taken from the simulation shown in Figure~\ref{fig:chap4nomarlkyn0}.}
\label{fig:compvnvy}
\end{center}
\end{figure}

Figure \ref{fig:chap4nomarlkyn0} displays the results of such simulation. The first frame for $v_{1n}$ shows the initial spatial distribution of the perturbation. Initially, $v_{1y}$, the velocity component associated to Alfv\'en waves, is zero. As time evolves, a non zero $v_{1y}$ component appears because of the coupling introduced by the three-dimensional propagation. The panels for $v_{1y}$ in Figure~\ref{fig:chap4nomarlkyn0} show that unlike $v_{1n}$ the excited transversal perturbations are not globally distributed in the potential arcade, but only at preferred locations, around a few magnetic surfaces. When the $v_{1n}$ and $v_{1y}$ signals are measured at one of those locations, $x=0$, $z/L=0.35$, it is seen that the amplitude related to the fast-like perturbation decreases in time, while the amplitude of the Alfv\'en-like component of the perturbation increases in time, see Figure~\ref{fig:compvnvy}. This is an indication of the wave energy transfer due to the resonant coupling of the excited fast normal mode to the Alfv\'enic solution around the excited magnetic surface. For long times a decrease in the amplitude of the velocity component, $v_{1y}$, can be appreciated and is attributed to numerical damping, for the reasons explained in \S~\ref{alfven_wave}.

\begin{figure}[!t]
\begin{center}
\includegraphics[width=0.49\textwidth]{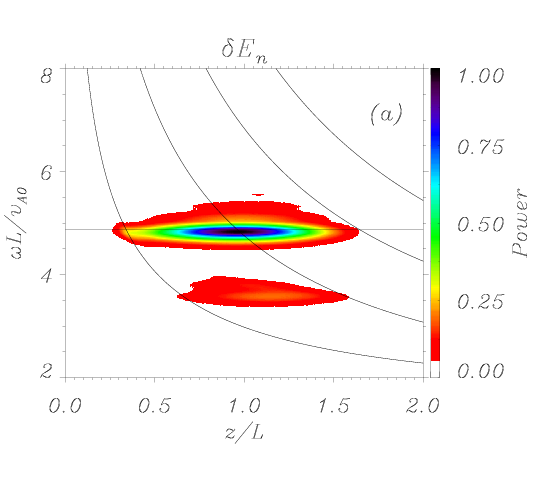}
\includegraphics[width=0.49\textwidth]{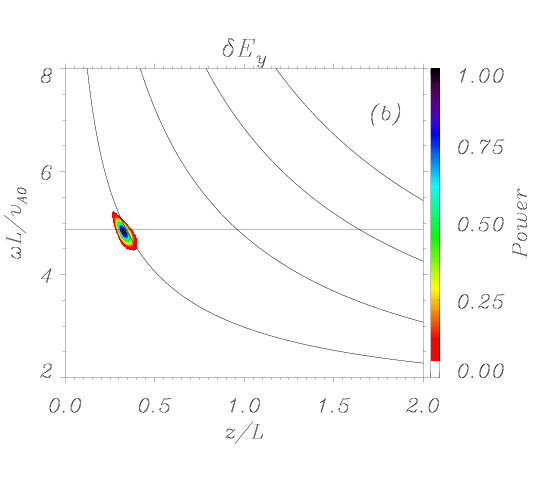}
\caption{The shaded contours represent the power spectrum of (a) $\delta E_n$ and (b) $\delta E_y$ as a function height, $z/L$, at the symmetry plane $x=0$, for the simulation shown in Figure \ref{fig:chap4nomarlkyn0}. Note that because of the quadratic nature of the wave energy density the curved lines have double the frequency of the Alfv\'en continua given by \citet{OBH1993} and the horizontal lines have  double the frequency of the fast normal mode.}
\label{fig:spectrumnormalky1}
\end{center}
\end{figure}

Further confirmation of the resonant wave energy transfer occurring between the modes can be obtained by computing the time evolution of the total energy density in our system. This total wave energy can be computed as

\begin{eqnarray}
\delta E(\mathbf{r},t)&=&\frac{1}{2}\left[\rho_{0}(v_{1x}^2+v_{1y}^2+v_{1z}^2)+\frac{1}{\mu_{0}}(B_{1x}^2+B_{1y}^2+B_{1z}^2)\right]. \label{eq:totalenergy}\\
\nonumber
\end{eqnarray}
The right-hand side panels in Figure~\ref{fig:chap4nomarlkyn0} show the spatial distribution of this quantity as a function of time. The different frames clearly indicate that, initially, the energy is distributed globally around the center of the system, such as corresponds to the initial perturbation we have used. At later times, this energy is transferred to magnetic surfaces around the particular magnetic field line in the arcade where the signals in Figure~\ref{fig:compvnvy} have been measured. The location of this energy deposition is not an arbitrary one. As previous theoretical works on the resonant energy transfer have shown, \citep[e.g., ][]{W1992,HG1993,RGB1997,AOB2004,RW2009}, global fast modes resonantly couple to localized Alfv\'en continuum modes at the magnetic surfaces where the frequency of the fast mode matches that of the corresponding Alfv\'en mode. In our case the spectral analysis of the wave energy densities associated to the normal and perpendicular components, plotted in Figure \ref{fig:spectrumnormalky1}, allow us to confirm the resonant energy transfer at the location where the fundamental fundamental fast mode frequency crosses the Alfv\'en continuum, that exactly corresponds to the magnetic surface where Alfv\'enic oscillations are excited and energy transfer occurs, see Figure \ref{fig:spectrumnormalky1}b. Although the fast mode frequency crosses other Alfv\'en continua, coupling can only occur if the parity of the fast and Alfv\'en eigenfunctions along the field lines is the same, see further details in \citet{AOB2004}. This prevents the coupling with Alfv\'en continuum modes with two extrema along field lines. Even if the coupling with Alfv\'en modes with three extrema along field lines is allowed, we find no signatures of this resonant coupling in the power spectrum analysis nor the wave energy density evolution.


	
\subsection{Propagation of coupled fast and Alfv\'en disturbances in a potential coronal arcade}
\label{gaussian_like}
Oscillations in coronal magnetic structures are believed to be generated by nearby disturbances, such as flares or filament eruptions. It is clear that such disturbances are far from being a normal mode of a particular structure as our potential arcade. Therefore, we have next considered the impulsive excitation of perturbations by means of a localized disturbance, which is expected to be a better representation of the real phenomena that often trigger waves and oscillations in the solar corona. In particular a Gaussian velocity perturbation is considered and the response of the system is expected to be different from the one described in \S~\ref{normal_like}, since now the initial perturbation is likely to be decomposed in a linear combination of normal modes with different frequencies that will constitute the resulting propagating wave.



\begin{figure*}[!t]
\begin{center}
\hspace{-0.3cm}\includegraphics[width=0.31\textwidth]{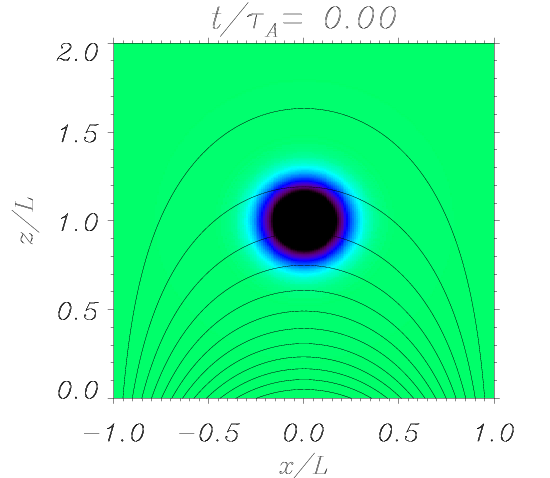}
\includegraphics[width=0.31\textwidth]{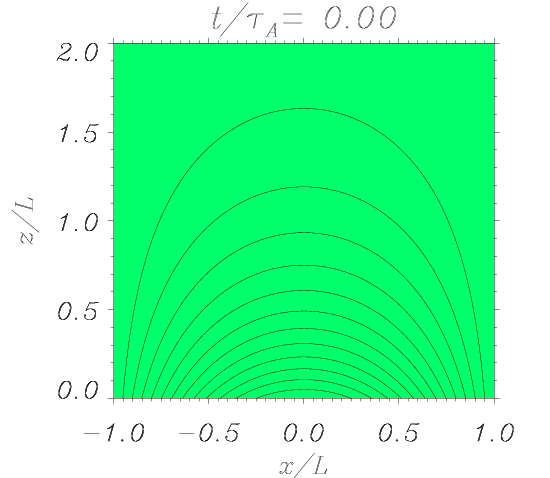}
\includegraphics[width=0.31\textwidth]{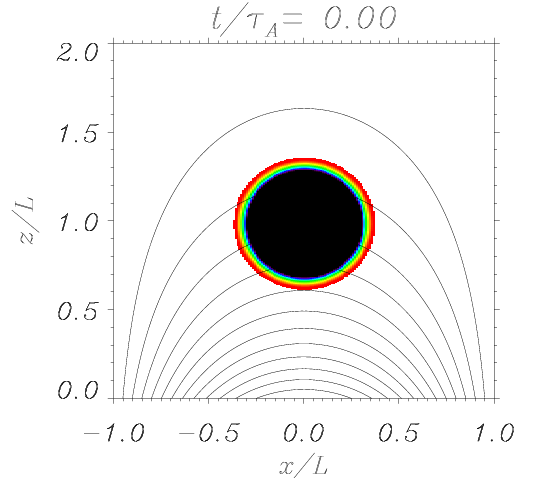}\\
\hspace{-0.3cm}\includegraphics[width=0.31\textwidth]{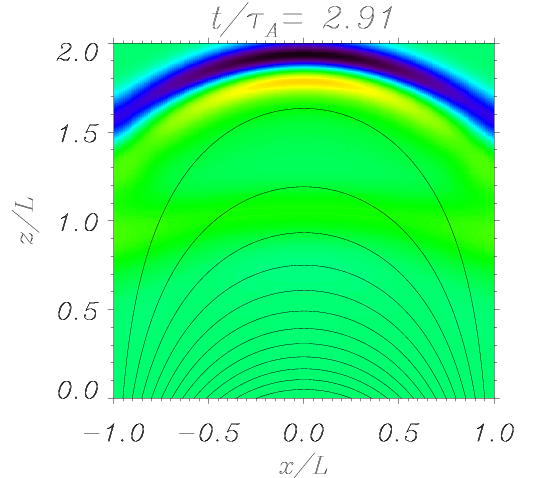}
\includegraphics[width=0.31\textwidth]{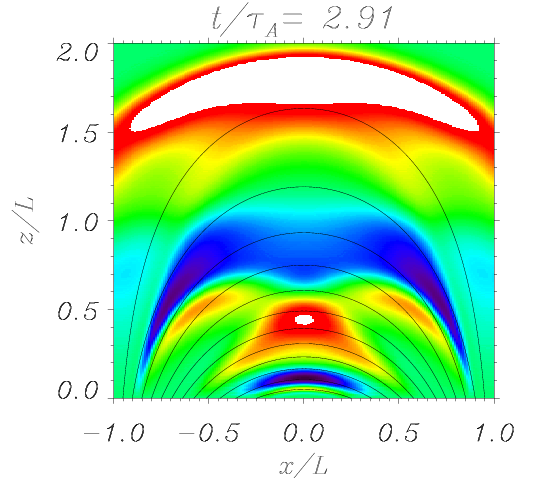}
\includegraphics[width=0.31\textwidth]{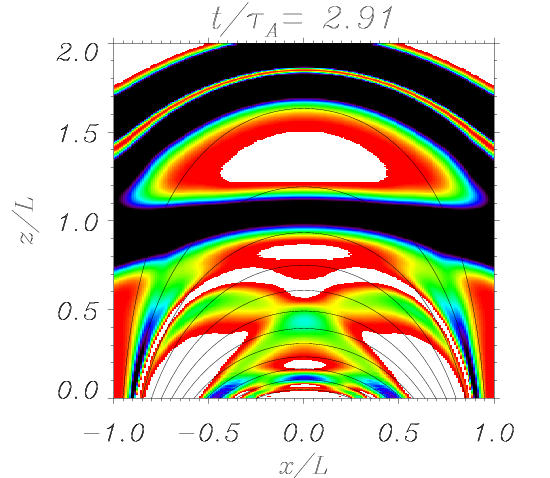}\\
\hspace{-0.3cm}\includegraphics[width=0.31\textwidth]{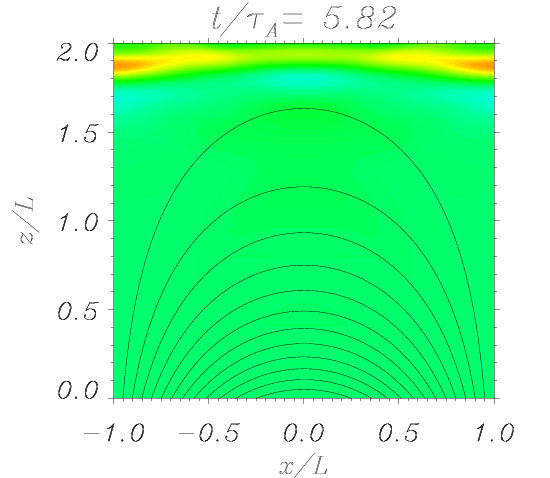}
\includegraphics[width=0.31\textwidth]{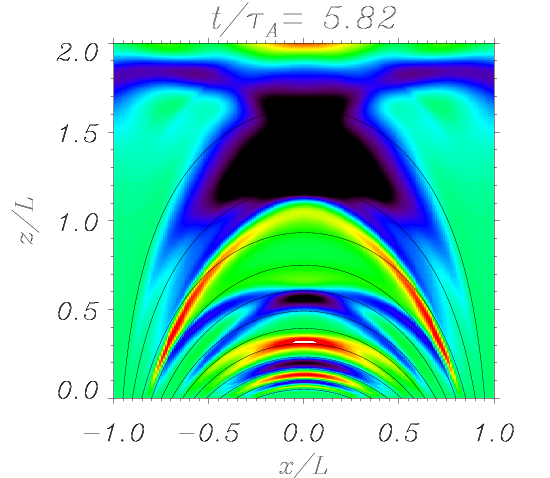}
\includegraphics[width=0.31\textwidth]{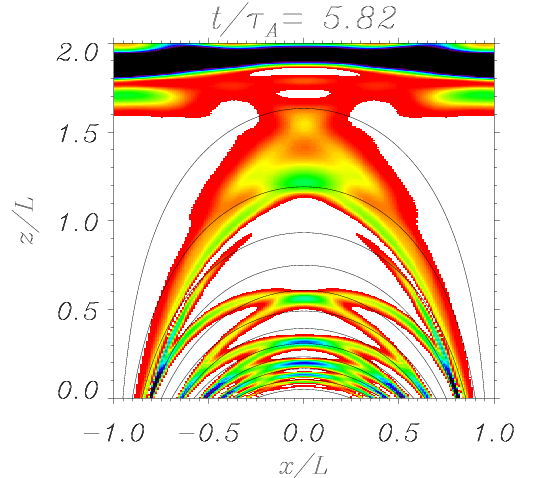}\\
\hspace{-0.3cm}\includegraphics[width=0.31\textwidth]{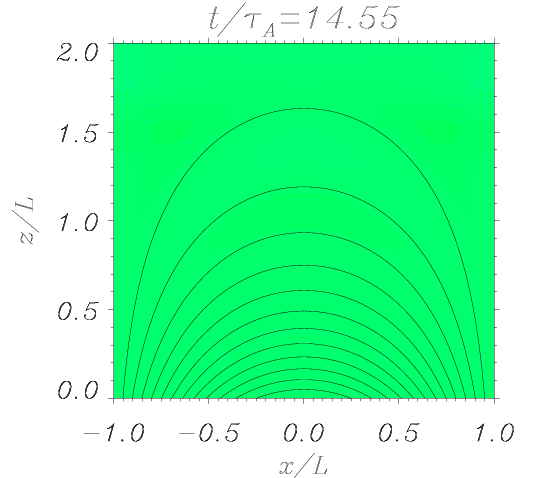}
\includegraphics[width=0.31\textwidth]{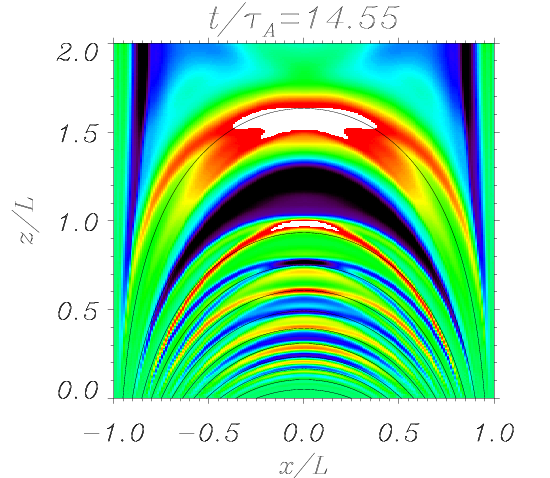}
\includegraphics[width=0.31\textwidth]{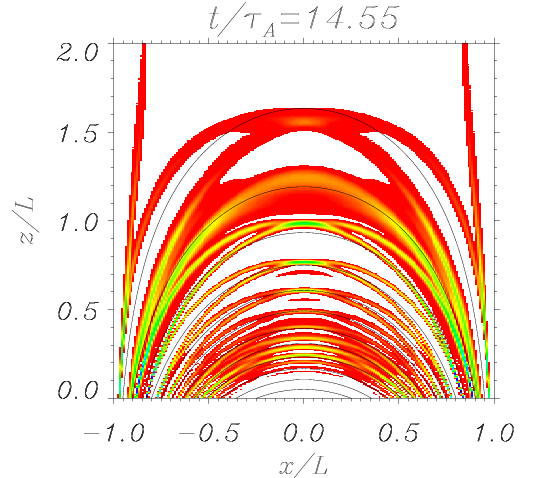}\\
\hspace{0.4cm}\includegraphics[width=0.31\textwidth]{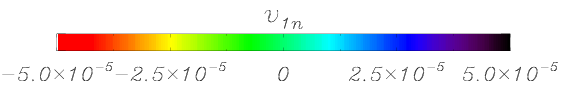}
\includegraphics[width=0.31\textwidth]{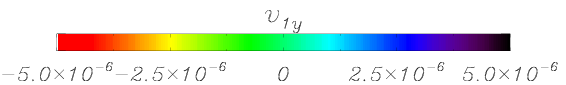}
\hspace{-0.1cm}\includegraphics[width=0.31\textwidth]{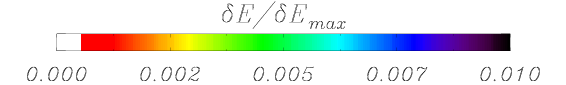}
\caption{Time evolution of the $v_{1n}$ (left panels), $v_{1y}$ (middle panels) velocity components, and of the total energy density (right panels) in a potential arcade with $\delta=1$. The initial perturbation is imposed on the $v_{1n}$ component with $x_s=0$, $z_s=1$, $v_s=10^{-4} v_{A0}$, $a=0.2L$ (see Equation~\ref{eq:perturbation}) and $k_{y}L=1$. For this simulation a $600\times600$ grid is used. Magnetic field lines are represented with white (left and middle panels) and black lines (right panels). A movie displaying the full time evolution is available in the electronic version of the journal.}
\label{fig:gaussd1kyn0}
\end{center}
\end{figure*}

We have produced an impulsive excitation of the $v_{1n}$ velocity component of the form given by Equation~(\ref{eq:perturbation}) and have considered $k_{y}\neq0$. Time evolution of the velocity components and the total energy density are displayed in Figure~\ref{fig:gaussd1kyn0}, which shows that the generated wave has both normal and perpendicular velocity components. Note that $v_{1y}=0$ in the absence of $k_{y}$ \citep[see][]{TOB2008}. It is clear in Figure~\ref{fig:gaussd1kyn0} that the perturbed normal velocity component evolution is similar to the one presented by \citet{TOB2008}, for the decreasing Alfv\'en speed model with constant density and $k_{y}=0$. For the normal velocity component, the shape of the wavefront is not circular, due to the fact that perturbations propagate faster toward the photosphere. For large times the front tends to be planar as the initial curvature of the wave packet is lost. As for the perpendicular velocity perturbation that is excited because of the three-dimensional character of the wave, its spatial distribution is highly anisotropic, with the signal concentrated around many magnetic surfaces. A wavefront with fast-like properties, similar to the one present in $v_{1n}$, can also be seen to propagate upwards producing $v_{1y}$ perturbations until it leaves the system. At the end, a collection of Alfv\'enic oscillations are generated in the arcade. By comparing with the results presented in the previous section we can think about them as being generated by the resonant coupling between the fast-like wavefront and several Alfv\'en continuum solutions, instead of the single resonance case shown in \S~\ref{normal_like}. Once excited, magnetic surfaces remain oscillating with their natural period and for large times they are phase-mixed because of the transverse non-uniformity.   

\begin{figure}[!t]
\begin{center}
\hspace{-0.3cm}\includegraphics[width=0.3\textwidth]{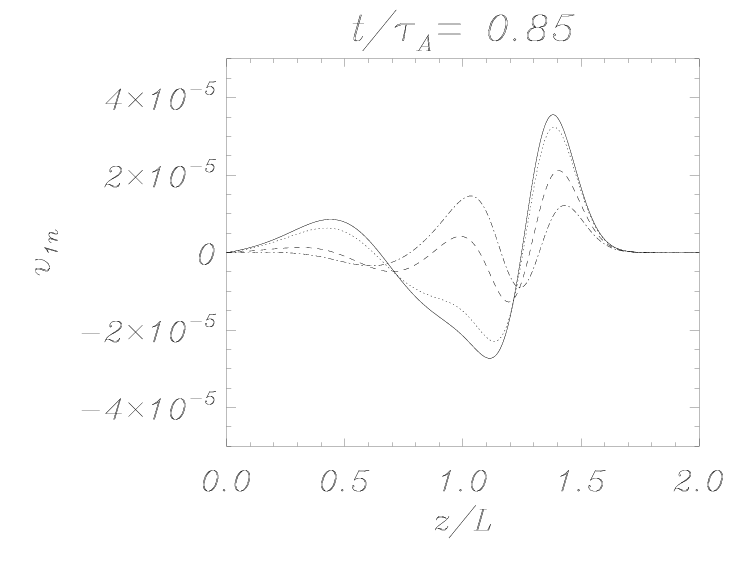}
\includegraphics[width=0.3\textwidth]{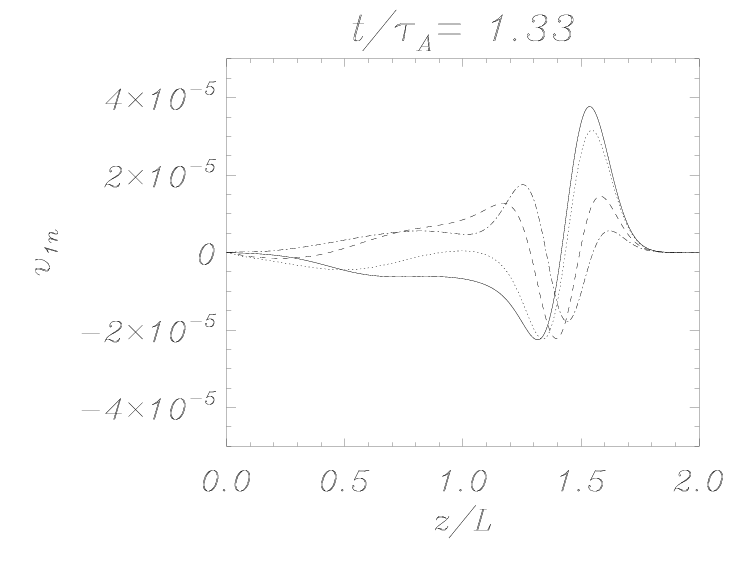}
\includegraphics[width=0.3\textwidth]{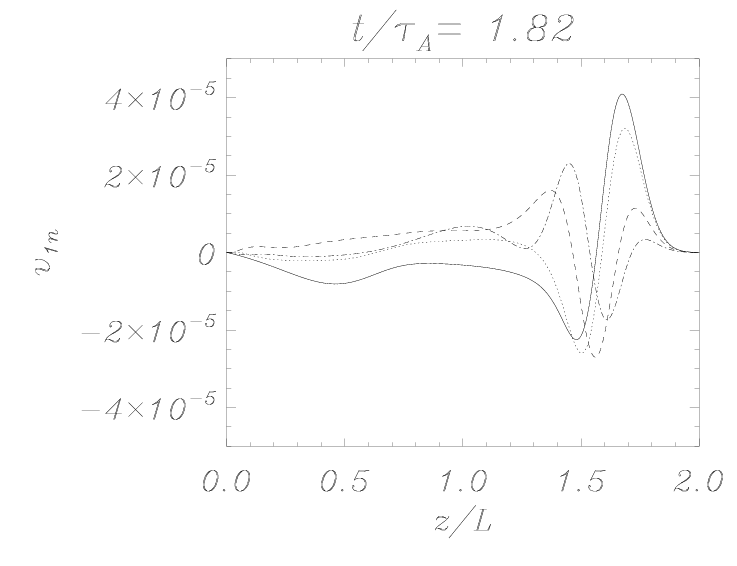}\\
\includegraphics[width=0.3\textwidth]{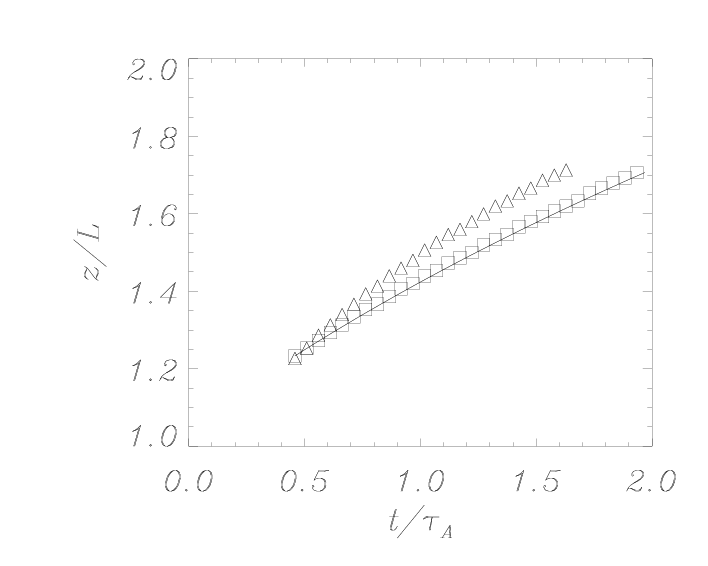}
\includegraphics[width=0.3\textwidth]{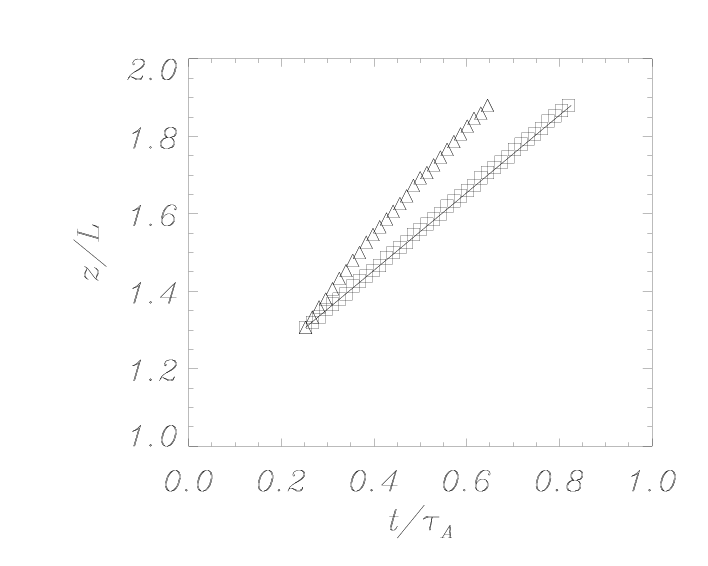}
\includegraphics[width=0.3\textwidth]{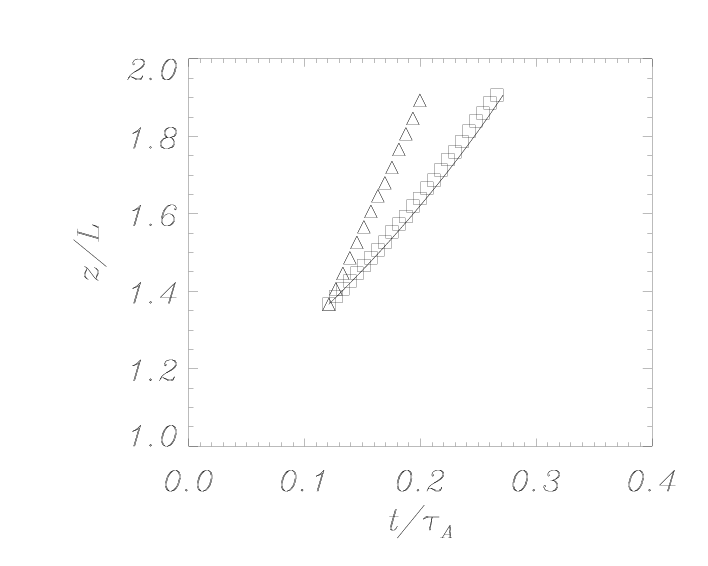}
\caption{\textbf{Top-panels} Several snapshots of the spatial distribution of the normal velocity component along $x=0$. The values of the longitudinal wavenumber are $k_{y}=0$ (solid line), $k_{y}L=3$ (dotted), $k_{y}L=7$ (dashed), and $k_{y}L=10$ (dash-dotted). \ \textbf{Bottom-panels} Position of the wavefront as a function of time when different values of the delta parameter, $\delta=1$ (left), $\delta=2$ (middle), and $\delta=3$ (right), and longitudinal wavenumber, $k_{y}=0$ (squares) and $k_{y}L=10$ (triangles), are selected. The solid line shows the analytical solution for the wavefront position when longitudinal propagation is not allowed ($k_{y}=0$); see Equation~(\ref{eq:localalfven}).}
\label{fig:posmaxkyn0}
\end{center}
\end{figure}

We have next analyzed in a quantitative way the effect of $k_{y}$ on the properties of the generated fast-like wavefront and the induced Alfv\'enic oscillations. Regarding the fast-like wavefront, Figure~\ref{fig:posmaxkyn0} (top-panels) shows different snapshots of the cut along $x=0$ of the $v_{1n}$ component for different values of $k_{y}$. These figures indicate that the larger the value of $k_{y}$ the faster the wavefront propagates. The propagation velocity can be measured by plotting the position of the wavefront maximum as a function of time (see Figure~\ref{fig:posmaxkyn0} bottom-left). The time evolution of the wavefront is followed for $t\geqslant t_{0}$ and the initial position of the maximum is denoted by $z_{0}$. For this relatively simple case, the numerical results can be compared with the analytical formula obtained by the integration of the local Alfv\'en speed profile (see Equation~[$5$] in Oliver et al. 1998). The resulting expression is

\begin{equation}
z(t)=\frac{2\Lambda_{B}}{2-\delta}\log\left[ \pm v_{A0}\frac{2-\delta}{2\Lambda_{B}}(t-t_{0})+\exp\left(\frac{2-\delta}{2\Lambda_{B}}z_{0} \right)  \right],
\label{eq:localalfven}
\end{equation}
where the $+$ and $-$ signs correspond to upward and downward propagation, respectively. Figure~\ref{fig:posmaxkyn0} (bottom-panels) shows a perfect correspondence between the numerically measured speed and the analytical expression when different models of the solar atmosphere are considered. The increase of the travel speed of fast-like wavefronts when $k_{y}\neq0$ is an important property to be taken into account in the three-dimensional problem.

\begin{figure}[t!]
\includegraphics[width=\textwidth]{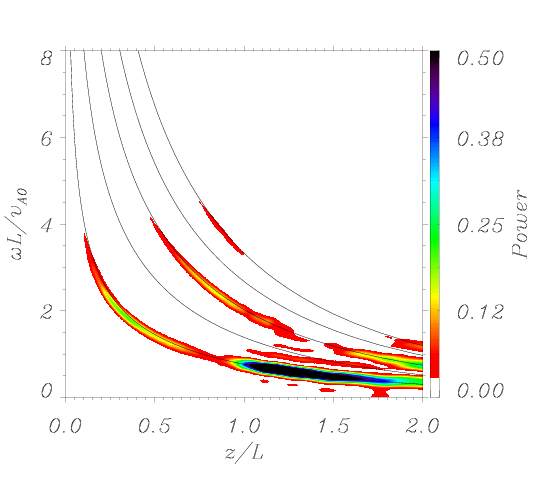}
\caption{Shaded contours represent the normalized power spectrum of the $v_{1y}$ velocity component corresponding to the simulation show in Figure~\ref{fig:gaussd1kyn0} as a function of the maximum height of field lines, $z/L$, and normalized frequency, $\omega L/v_{A0}$. Solid lines are the theoretical frequency of the Alfv\'en normal mode obtained by \citet{OBH1993}. The frequency analysis is made at the symmetry plane, $x=0$.}
\label{fig:spectrumd1ky1}
\end{figure}

For the Alfv\'enic oscillations the power spectrum is analyzed in a cut along $x=0$, which allows us to study the power on different magnetic surfaces. Figure \ref{fig:spectrumd1ky1} shows power at a large number of magnetic surfaces, not just around a selected group of field lines around a given magnetic surface, so a wide range of magnetic surfaces are excited because of the coupling. Also, not just the fundamental mode is excited, but also several higher harmonics. All of them have even parity with respect to $x=0$, such as corresponds to the parity of the $v_{1n}$ perturbation and the parity rule for $k_{y}\neq0$ \citep{AOB2004}. When comparing the power spectrum obtained from the numerical solution with the analytical Alfv\'en continuum frequencies given by \citet{OBH1993} for the case $k_{y}=0$, we see that the signal coincides with the analytical curves for $k_{y}=0$. Such as expected,  perpendicular propagation has no effect on the frequencies of Alfv\'en waves generated on different magnetic surfaces in the arcade. This is a known result since $\omega\sim(\mathbf{k}\cdot\mathbf{B})$.

\begin{figure}[!t]
\begin{center}
\includegraphics[width=\textwidth]{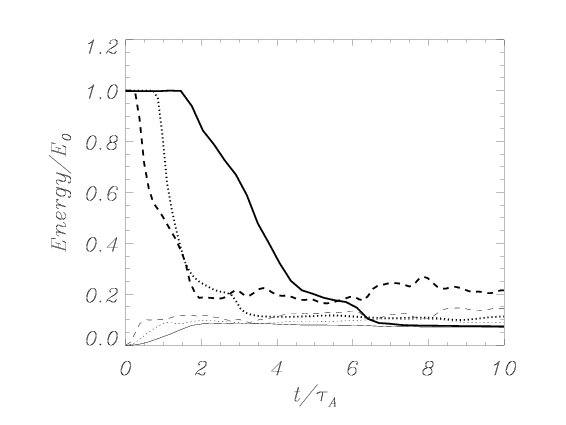}
\caption{The solid ($\delta=1$), dotted ($\delta=2$), and dashed ($\delta=3$)  lines are the normalized total energy (thick lines) and the normalized total energy associated to the $y$-direction (thin lines) as a function of time for $k_{y}L=1$.}
\label{fig:energiesdelta}
\end{center}
\end{figure}

As with the normal mode case, a quantitative analysis of the time-evolution of the wave energy of the system helps to better understand the process of energy conversion between fast and Alfv\'en waves. Figure~\ref{fig:gaussd1kyn0} (right-hand side panels), shows the evolution of the total energy density as a function of time. At early stages this quantity shows a clear signature of a fast-like wavefront propagating through the domain. For long times the energy deposition is spatially distributed on the whole system, not only on a single magnetic surface, as in the previous section. Although a large part of the energy leaves the system in the form of fast-like wavefronts, part of the energy remains trapped in the Alfv\'enic oscillations that are resonantly excited in the arcade. The amount of energy trapped in the system can be calculated by the integration of the total energy density (see Equation~[\ref{eq:totalenergy}]) in the whole domain as a function of time. The result is shown in Figure~\ref{fig:energiesdelta} (solid line). For short times the total energy remains almost constant, but when the fast front reaches the boundaries of the system a strong decrease of this quantity is seen. Resonant wave conversion very quickly produces velocity perturbations in the $y$-direction and the energy associated to these Alfv\'enic components grows up to its maximum value before the fast wavefront leaves the system. At later stages, a fraction of around $5-10 \%$ of the initial total energy is retained in the system and the total energy remains almost constant in the subsequent time evolution. We must note that this energy is trapped even in the absence of any density enhancement or wave cavity.

\begin{figure}[!t]
\begin{center}
\includegraphics[width=0.46\textwidth]{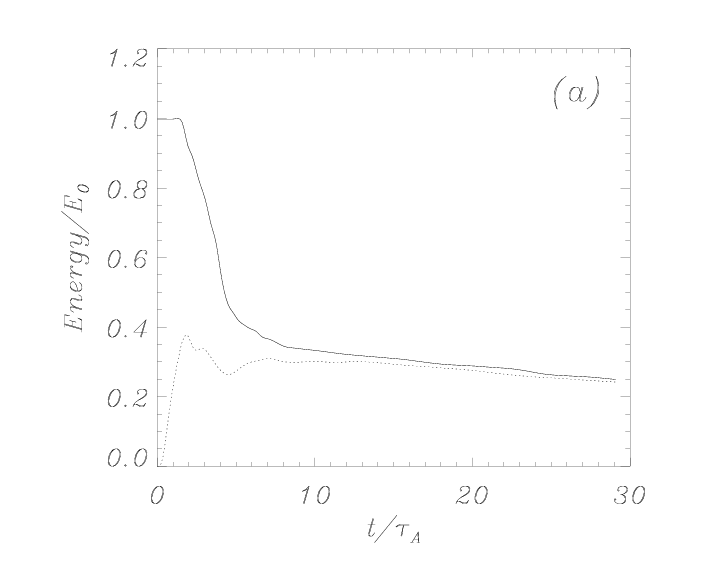}
\includegraphics[width=0.46\textwidth]{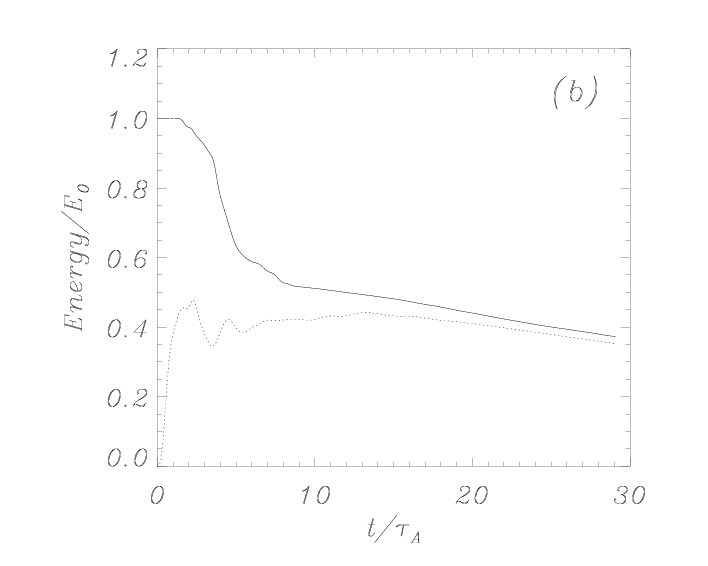}
\includegraphics[width=0.46\textwidth]{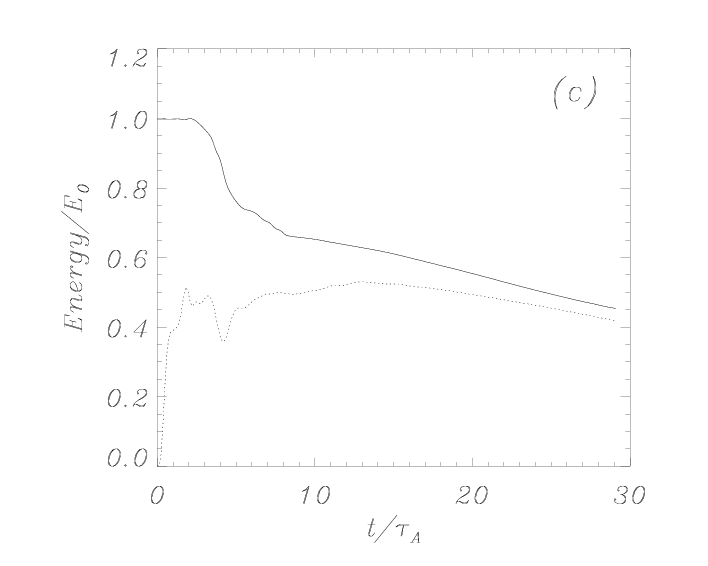}
\includegraphics[width=0.46\textwidth]{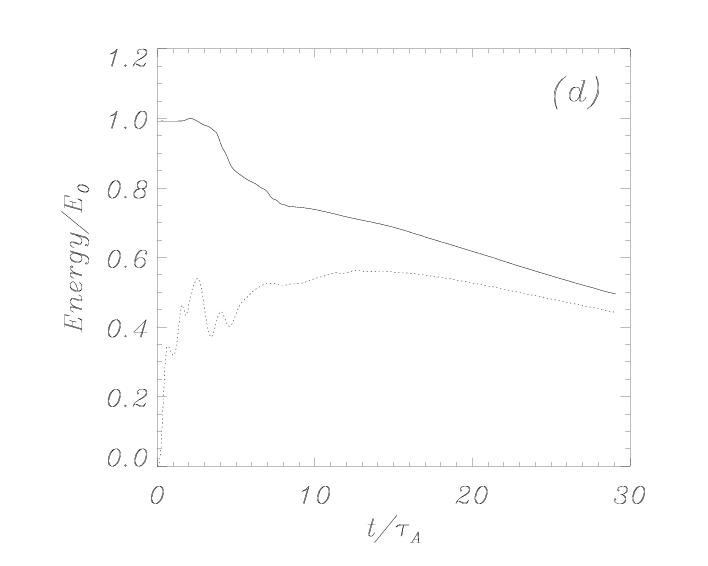}
\caption{Normalized total energy of the system as a function of time (solid line) and normalized energy associated to the $y$-direction (dotted line) for $\delta=1$ and different values of the longitudinal wavenumber. (a) $k_{y}L=3$, (b) $k_{y}L=5$, (c) $k_{y}L=7$, and (d) $k_{y}L=9$.}
\label{fig:energiesky}
\end{center}
\end{figure}

So far, we have used fixed values for the perpendicular propagation wavenumber, $k_{y}$, and the ratio of magnetic to density scale heights, $\delta$. We have next analyzed the influence of these parameters on the obtained results concerning the energy transfer between fast and Alfv\'en waves. Figure~\ref{fig:energiesky} shows the total energy time evolution for different values of $k_{y}$. Several conclusions can be extracted. First, the amount of energy that is trapped by the system in the form of Alfv\'enic oscillations increases with $k_{y}$ and is above $40\%$ for the largest value of this parameter that we have considered. This can be understood in terms of stronger resonant coupling occurring for larger values of $k_{y}$. The relation between the total energy and the energy associated to the $y$-direction also changes with $k_{y}$, in such a way that, while for relatively small $k_{y}$ almost all the energy of the system is stored in oscillations in the $y$-direction, for larger values of $k_{y}$ there is a difference between the total energy and the Alfv\'enic energy for large times. To understand this we need to mention that for $k_{y}\neq0$ Alfv\'en waves have both perpendicular and normal velocity components and so Alfv\'en wave energy is not only contained in the $y$-direction. Although this effect is less visible in the simulations it can be measured, such as shown in Figure~\ref{fig:energiesky}. Note also that for large times the two energy densities decay. This is due to numerical damping, since when very small scales are created the spatial resolution used is not fine enough to handle the localized Alfv\'enic oscillations that are phase-mixed for large times (see \S~\ref{alfven_wave}).


As for the $\delta$ parameter, it controls our model atmosphere, since it allows us to select different ratios of the magnetic scale height to the density scale height. By repeating the previous numerical experiments for two additional values of this parameter the following results are obtained (see Figure~\ref{fig:posmaxkyn0} bottom-panels). Depending on the value of $\delta$, the Alfv\'en speed profile in the vertical direction has a steeper or flatter profile. This means that the time that a fast-like perturbation needs to reach the boundaries of the system and leave it varies with $\delta$. Therefore, the time at which the sudden decrease of the total energy of the system occurs differs for different values of $\delta$, see Figure~\ref{fig:energiesdelta}. However, the fractional amount of wave energy that is transferred to Alfv\'en waves and is trapped in the system does not depend on the model atmosphere we use. Nevertheless, the rate at which the energy transfer occurs does depend in the model atmosphere, such as can be appreciated from the different slopes of the energy in Figure~\ref{fig:energiesdelta}.

\section{Conclusions}
\label{conclusions}
In this paper we have studied the temporal evolution of coupled fast and Alfv\'en waves in a potential coronal arcade when three-dimensional propagation is allowed. Because of the inclusion of three-dimensional dependence on the perturbed quantities, fast and Alfv\'en waves are coupled and the resulting solutions display a mixed fast/Alfv\'en character. The non-uniform nature of the considered medium produces the coupling to be of resonant nature, in such a way that transfer of energy and wave damping occur in the system.

First, the nature of resonant coupling between a fast normal mode of the system and Alfv\'en continuum modes has been analyzed. It is seen that the fast mode with a global nature resonantly couples to localized Alfv\'en waves around a given magnetic surface in the arcade, thus transferring its energy to the later. The position of the resonant surface perfectly agrees with the resonant frequency condition predicted by several authors in previous studies of this kind, and with the parity rules given by \citet{AOB2004}.

Next, the temporal evolution of a localized impulsive disturbance has been analyzed. The inclusion of perpendicular propagation produces an increase in the wave propagation speed for the fast-like wavefront when compared to the purely poloidal propagation case. As in the previous case, perpendicular propagation induces the excitation of Alfv\'enic oscillations around magnetic surfaces, due to the resonant coupling between fast and Alfv\'en waves. Now these oscillations cover almost the whole domain in the arcade, so that the energy of the initial perturbation is spread into localized Alfv\'enic waves. The frequency of the induced Alfv\'enic oscillations is seen to be independent from the perpendicular wavenumber. As time progresses and the initial wavefront leaves the system part of the energy is stored in these Alfv\'en waves which remain confined around magnetic surfaces. Phase mixing then gives rise to smaller and smaller spatial scales, until the numerical code is unable to properly follow the subsequent time evolution. The energy trapping around magnetic surfaces occurs even in the absence of a density enhancement or a wave cavity structure, and is only due to the non-uniformity of the density profile and the magnetic structuring, which lead to a non-uniform Alfv\'en speed distribution.

Finally, the efficiency of this wave energy transfer between large scale disturbances and small scale oscillations has been studied as a function of the perpendicular wavenumber and for different values of the ratio of the magnetic scale height to the density scale height. It is seen that the first factor strongly affects the amount of energy trapped by Alfv\'en waves. The amount of energy trapped by the arcade increases for increasing value of the perpendicular wavenumber. The particular ratio of magnetic to density scale heights determines how fast the available fast wave energy leaves the system and, therefore, the rate at which energy can be transferred to Alfv\'en waves, but not the final amount of energy stored by the arcade in the form of Alfv\'enic oscillations. These $2.5$D simulations should be extended in the future to more realistic $3$D simulations in order to ascertain the applicability of our conclusions to the real wave dynamics observed in coronal structures.

\acknowledgments

The authors acknowledge the Spanish MCyT for the funding provided under project AYA$2006-07637$ and D. Fanning (http://www.dfanning.com/) for his helpful advices about IDL. S.R. also acknowledges MCyT for a fellowship.


\begin{thebibliography}{41}
\expandafter\ifx\csname natexlab\endcsname\relax\def\natexlab#1{#1}\fi

\bibitem[{{Arregui} {et~al.}(2004{\natexlab{a}}){Arregui}, {Oliver}, \&
  {Ballester}}]{AOB2004}
{Arregui}, I., {Oliver}, R., \& {Ballester}, J.~L. 2004{\natexlab{a}}, \aap,
  425, 729

\bibitem[{{Arregui} {et~al.}(2004{\natexlab{b}}){Arregui}, {Oliver}, \&
  {Ballester}}]{AOB2004a}
---. 2004{\natexlab{b}}, \apj, 602, 1006

\bibitem[{{Aschwanden} {et~al.}(2002){Aschwanden}, {de Pontieu}, {Schrijver},
  \& {Title}}]{APS2002}
{Aschwanden}, M.~J., {de Pontieu}, B., {Schrijver}, C.~J., \& {Title}, A.~M.
  2002, \solphys, 206, 99

\bibitem[{{Aschwanden} {et~al.}(1999){Aschwanden}, {Fletcher}, {Schrijver}, \&
  {Alexander}}]{AFS1999}
{Aschwanden}, M.~J., {Fletcher}, L., {Schrijver}, C.~J., \& {Alexander}, D.
  1999, \apj, 520, 880

\bibitem[{{Banerjee} {et~al.}(2007){Banerjee}, {Erd{\'e}lyi}, {Oliver}, \&
  {O'Shea}}]{BEO2007}
{Banerjee}, D., {Erd{\'e}lyi}, R., {Oliver}, R., \& {O'Shea}, E. 2007,
  \solphys, 246, 3

\bibitem[{{Bona} {et~al.}(2009){Bona}, {Bona-Casas}, \& {Terradas}}]{BBT2009}
{Bona}, C., {Bona-Casas}, C., \& {Terradas}, J. 2009, Journal of Computational
  Physics, 228, 2266

\bibitem[{{Brady} \& {Arber}(2005)}]{BA2005}
{Brady}, C.~S., \& {Arber}, T.~D. 2005, \aap, 438, 733

\bibitem[{{Brady} {et~al.}(2006){Brady}, {Verwichte}, \& {Arber}}]{BVA2006}
{Brady}, C.~S., {Verwichte}, E., \& {Arber}, T.~D. 2006, \aap, 449, 389

\bibitem[{{{\v C}ade{\v z} } \& {Ballester}(1995{\natexlab{a}})}]{CB1995}
{{\v C}ade{\v z} }, V.~M., \& {Ballester}, J.~L. 1995{\natexlab{a}}, \aap, 296,
  537

\bibitem[{{{\v C}ade{\v z} } \& {Ballester}(1995{\natexlab{b}})}]{CB1995a}
---. 1995{\natexlab{b}}, \aap, 296, 550

\bibitem[{{Del Zanna} {et~al.}(2005){Del Zanna}, {Schaekens}, \&
  {Velli}}]{DSV2005}
{Del Zanna}, L., {Schaekens}, E., \& {Velli}, M. 2005, \aap, 431, 1095

\bibitem[{{D{\'{\i}}az} {et~al.}(2006){D{\'{\i}}az}, {Zaqarashvili}, \&
  {Roberts}}]{DAR2006}
{D{\'{\i}}az}, A.~J., {Zaqarashvili}, T., \& {Roberts}, B. 2006, \aap, 455, 709

\bibitem[{{Edwin} \& {Roberts}(1983)}]{ER1983}
{Edwin}, P.~M., \& {Roberts}, B. 1983, \solphys, 88, 179

\bibitem[{{Goossens} {et~al.}(1985){Goossens}, {Poedts}, \&
  {Hermans}}]{GPH1985}
{Goossens}, M., {Poedts}, S., \& {Hermans}, D. 1985, \solphys, 102, 51

\bibitem[{{Halberstadt} \& {Goedbloed}(1993)}]{HG1993}
{Halberstadt}, G., \& {Goedbloed}, J.~P. 1993, \aap, 280, 647

\bibitem[{{Murawski} {et~al.}(2005){Murawski}, {Selwa}, \& {Nocera}}]{MSN2005}
{Murawski}, K., {Selwa}, M., \& {Nocera}, L. 2005, \aap, 437, 687

\bibitem[{{Nakariakov} \& {Ofman}(2001)}]{NO2001}
{Nakariakov}, V.~M., \& {Ofman}, L. 2001, \aap, 372, L53

\bibitem[{{Nakariakov} {et~al.}(1999){Nakariakov}, {Ofman}, {Deluca},
  {Roberts}, \& {Davila}}]{NOD1999}
{Nakariakov}, V.~M., {Ofman}, L., {Deluca}, E.~E., {Roberts}, B., \& {Davila},
  J.~M. 1999, Science, 285, 862

\bibitem[{{Oliver} {et~al.}(1993){Oliver}, {Ballester}, {Hood}, \&
  {Priest}}]{OBH1993}
{Oliver}, R., {Ballester}, J.~L., {Hood}, A.~W., \& {Priest}, E.~R. 1993, \aap,
  273, 647

\bibitem[{{Oliver} {et~al.}(1996){Oliver}, {Hood}, \& {Priest}}]{OHP1996}
{Oliver}, R., {Hood}, A.~W., \& {Priest}, E.~R. 1996, \apj, 461, 424

\bibitem[{{Oliver} {et~al.}(1998){Oliver}, {Murawski}, \&
  {Ballester}}]{OMB1998}
{Oliver}, R., {Murawski}, K., \& {Ballester}, J.~L. 1998, \aap, 330, 726

\bibitem[{{Poedts} \& {Goossens}(1988)}]{PG1988}
{Poedts}, S., \& {Goossens}, M. 1988, \aap, 198, 331

\bibitem[{{Poedts} {et~al.}(1985){Poedts}, {Hermans}, \& {Goossens}}]{PHG1985}
{Poedts}, S., {Hermans}, D., \& {Goossens}, M. 1985, \aap, 151, 16

\bibitem[{{Roberts} {et~al.}(1984){Roberts}, {Edwin}, \& {Benz}}]{REB1984}
{Roberts}, B., {Edwin}, P.~M., \& {Benz}, A.~O. 1984, \apj, 279, 857

\bibitem[{{Ruderman} {et~al.}(1997){Ruderman}, {Goossens}, {Ballester}, \&
  {Oliver}}]{RGB1997}
{Ruderman}, M.~S., {Goossens}, M., {Ballester}, J.~L., \& {Oliver}, R. 1997,
  \aap, 328, 361

\bibitem[{{Russell} \& {Wright}(2009)}]{RW2009}
{Russell}, A. J.~B., \& {Wright}, A.~N. 2009, in press

\bibitem[{{Schrijver} {et~al.}(2002){Schrijver}, {Aschwanden}, \&
  {Title}}]{SAT2002}
{Schrijver}, C.~J., {Aschwanden}, M.~J., \& {Title}, A.~M. 2002, \solphys, 206,
  69

\bibitem[{{Selwa} {et~al.}(2007){Selwa}, {Murawski}, {Solanki}, \&
  {Wang}}]{SMS2007}
{Selwa}, M., {Murawski}, K., {Solanki}, S.~K., \& {Wang}, T.~J. 2007, \aap,
  462, 1127

\bibitem[{{Selwa} {et~al.}(2006){Selwa}, {Solanki}, {Murawski}, {Wang}, \&
  {Shumlak}}]{SSM2006}
{Selwa}, M., {Solanki}, S.~K., {Murawski}, K., {Wang}, T.~J., \& {Shumlak}, U.
  2006, \aap, 454, 653

\bibitem[{{Terradas}(2009)}]{T2009}
{Terradas}, J. 2009, Space Science Reviews, in press. doi:10.1007/s11214-009-9560-3 

\bibitem[{{Terradas} {et~al.}(1999){Terradas}, {Oliver}, \&
  {Ballester}}]{TOB1999}
{Terradas}, J., {Oliver}, R., \& {Ballester}, J.~L. 1999, \apj, 517, 488

\bibitem[{{Terradas} {et~al.}(2006){Terradas}, {Oliver}, \&
  {Ballester}}]{TOB2006}
---. 2006, \apjl, 650, L91

\bibitem[{{Terradas} {et~al.}(2008){Terradas}, {Oliver}, {Ballester}, \&
  {Keppens}}]{TOB2008}
{Terradas}, J., {Oliver}, R., {Ballester}, J.~L., \& {Keppens}, R. 2008, \apj,
  675, 875

\bibitem[{{Tsubaki}(1988)}]{T1988}
{Tsubaki}, T. 1988, in Solar and Stellar Coronal Structure and Dynamics, ed.
  R.~C. {Altrock}, 140--149


\bibitem[{{Van Doorsselaere} {et~al.}(2004){Van Doorsselaere}, {Debosscher},
  {Andries}, \& {Poedts}}]{VDA2004}
{Van Doorsselaere}, T., {Debosscher}, A., {Andries}, J., \& {Poedts}, S. 2004,
  \aap, 424, 1065

\bibitem[{{Van Doorsselaere} {et~al.}(2009){Van Doorsselaere}, {Verwichte}, \&
  {Terradas}}]{VVT2009}
{Van Doorsselaere}, T., {Verwichte}, E., \& {Terradas}, J. 2009, Space Science
  Reviews, in press. doi:10.1007/s11214-009-9530-9

\bibitem[{{Verwichte} {et~al.}(2006{\natexlab{a}}){Verwichte}, {Foullon}, \&
  {Nakariakov}}]{VFN2006a}
{Verwichte}, E., {Foullon}, C., \& {Nakariakov}, V.~M. 2006{\natexlab{a}},
  \aap, 446, 1139

\bibitem[{{Verwichte} {et~al.}(2006{\natexlab{b}}){Verwichte}, {Foullon}, \&
  {Nakariakov}}]{VFN2006b}
---. 2006{\natexlab{b}}, \aap, 449, 769

\bibitem[{{Verwichte} {et~al.}(2006{\natexlab{c}}){Verwichte}, {Foullon}, \&
  {Nakariakov}}]{VFN2006c}
---. 2006{\natexlab{c}}, \aap, 452, 615

\bibitem[{{Verwichte} {et~al.}(2004){Verwichte}, {Nakariakov}, {Ofman}, \&
  {Deluca}}]{VNO2004}
{Verwichte}, E., {Nakariakov}, V.~M., {Ofman}, L., \& {Deluca}, E.~E. 2004,
  \solphys, 223, 77

\bibitem[{{Wright}(1992)}]{W1992}
{Wright}, A.~N. 1992, \jgr, 97, 6429

\end{thebibliography}

\clearpage

\clearpage

\clearpage

\clearpage

\clearpage

\clearpage

\clearpage

\clearpage

\clearpage

\clearpage

\clearpage

\end{document}